\documentclass[aps,prb,twocolumn,superscriptaddress,amsmath,amssymb,tightenlines,footinbib,10pt]{revtex4-1}

\usepackage{times}
\usepackage[dvipdfmx]{graphicx} 
\usepackage{bmpsize}
\usepackage[latin1]{inputenc}
\usepackage{mathrsfs}
\usepackage[dvipsnames,usenames]{color}
\usepackage{soul}
\usepackage[colorlinks,bookmarks=true,citecolor=blue,linkcolor=blue,urlcolor=blue, breaklinks=true]{hyperref}
\DeclareMathOperator{\Tr}{Tr}
\usepackage[percent]{overpic}

\usepackage{amssymb}
\usepackage{latexsym}
\usepackage{amsmath}

\usepackage{xcolor}
\usepackage{graphicx}
\usepackage{amsfonts}
\usepackage{tabularx}
\usepackage{stmaryrd}
\usepackage{bbold}
\usepackage{array}
\usepackage{mathtools}

\newcommand*{\citen}{}% generate error, if `\citen` is already in use
\DeclareRobustCommand*{\citen}[1]{%
  \begingroup
    \romannumeral-`\x % remove space at the beginning of \setcitestyle
    \setcitestyle{numbers}%
    \cite{#1}%
  \endgroup
}

\begin{document}

\title{Entanglement topological invariants for one-dimensional topological superconductors}
\author{P. Fromholz$^*$} 
\affiliation{The Abdus Salam International Centre for Theoretical Physics, Strada Costiera 11, 34151 Trieste, Italy}
\affiliation{SISSA, via Bonomea 265, 34136 Trieste, Italy}
\author{G. Magnifico$^*$} 
\affiliation{The Abdus Salam International Centre for Theoretical Physics, Strada Costiera 11, 34151 Trieste, Italy}
\affiliation{Dipartimento di Fisica e Astronomia dell'Universit\`a di Bologna, I-40127 Bologna, Italy}
\affiliation{INFN, Sezione di Bologna, I-40127 Bologna, Italy}
\author{V. Vitale}
\affiliation{The Abdus Salam International Centre for Theoretical Physics, Strada Costiera 11, 34151 Trieste, Italy}
\affiliation{SISSA, via Bonomea 265, 34136 Trieste, Italy}
\author{T. Mendes-Santos}
\affiliation{The Abdus Salam International Centre for Theoretical Physics, Strada Costiera 11, 34151 Trieste, Italy}
\affiliation{SISSA, via Bonomea 265, 34136 Trieste, Italy}
\author{M. Dalmonte} 
\affiliation{The Abdus Salam International Centre for Theoretical Physics, Strada Costiera 11, 34151 Trieste, Italy}
\affiliation{SISSA, via Bonomea 265, 34136 Trieste, Italy}
\date{\today}
%%%%%%%%%%%%%%%%%%%%% ABSTRACT %%%%%%%%%%%%%%%%%%%%%%%%%%%%%%%%%%
\begin{abstract}
Entanglement provides characterizing features of true topological order in two-dimensional systems. We show how entanglement of disconnected partitions defines topological invariants for one-dimensional topological superconductors. These order parameters quantitatively capture the entanglement that is possible to distill from the ground state manifold and are thus quantized to 0 or $\log 2$. Their robust quantization property is inferred from the underlying lattice gauge theory description of topological superconductors and is corroborated via exact solutions and numerical simulations. Transitions between topologically trivial and non-trivial phases are accompanied by scaling behavior, a hallmark of genuine order parameters, captured by entanglement critical exponents. These order parameters are experimentally measurable utilizing state-of-the-art techniques. 
\end{abstract}
%%%%%%%%%%%%%%%%%%%%%%%%%%%%%%%%%%%%%%%%%%%%%%%%%%%%%%%%%%%%%
 
\maketitle
%%%%%%%%%%%%%%%%%%%%%%%%%%%%%%%%%%%%%%%%%%%%%%%%%%%%%%%%%%%%%

\section{Introduction} 
Recently, entanglement has emerged as a groundbreaking diagnostic to characterize and classify many-body quantum phenomena in- and out-of-equilibrium~\cite{Amico2008,Calabrese_2009,Eisert2010,fradkinbook}. An archetypal example is the possibility of unambiguously detecting topological order in two-dimensional systems via the topological entanglement entropy (TEE)~\cite{Hamma_2005,2006PhRvL..96k0405L,Kitaev_2006}. The latter spots the presence of `long-range' entanglement which is not distillable via local operations; consequently, it defines a genuine entanglement order parameter, that distinguishes phases depending on their quasiparticle content~\cite{Wen19}. This insight has been widely employed in the characterization of topologically ordered states in numerical simulations~\cite{Depenbrock:2012aa,Jiang:2012aa,Isakov:2011aa}, and has stimulated the search for experimentally realistic entanglement probes~\cite{Abanin:2012aa,Daley_2012,Elben_2018,Vermersch_2018,Pichler:2016aa,Dalmonte:2017aa,Cornfeld_2019}. 

While the definition of the TEE naturally emerges from gauge theories in two-dimensions, the existence of topological invariants based solely on entanglement properties in one-dimensional (1D) topological matter - e.g., in the form of an order parameter - is presently not clear. In 1D, bipartite entanglement of \textit{connected} partitions does not display informative scaling corrections~\cite{Calabrese_2009,Eisert2010}. Its finer structure - captured by the entanglement spectrum -, provides sharp \textit{sine qua non}~\cite{Pollmann:2010aa,Fidkowski:aa,Turner:2011aa}, but even the entanglement spectrum of single partitions is not able to distinguish the topological character of wave functions~\footnote{Examples include the equivalence between the entanglement spectra of the ground state of finite Ising and Kitaev chains, and spin ladders~\cite{Pollmann:2010aa}.}. Indeed, at the field theory level and in 1D, connected bipartite entanglement is strongly influenced by ultra-violet contributions due to edges, and is thus not immediately linked to 'universal' information. 

\begin{figure}[h]
   \centering
   \begin{overpic}[width=1.\linewidth]{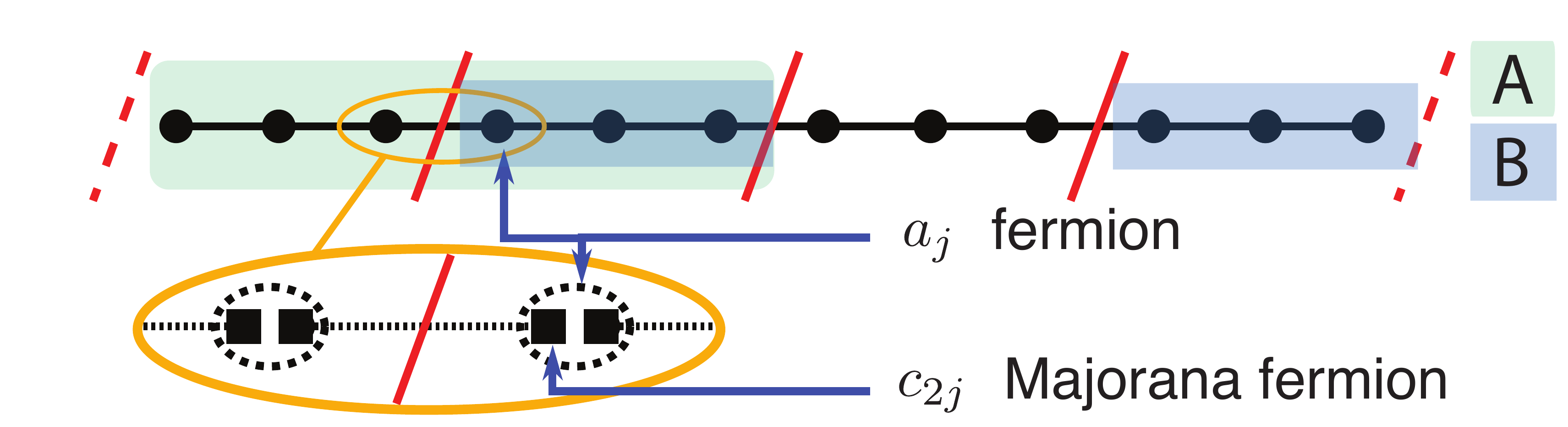}
    \put (0,10) {a)}
\end{overpic}
\begin{overpic}[width=1.\linewidth]{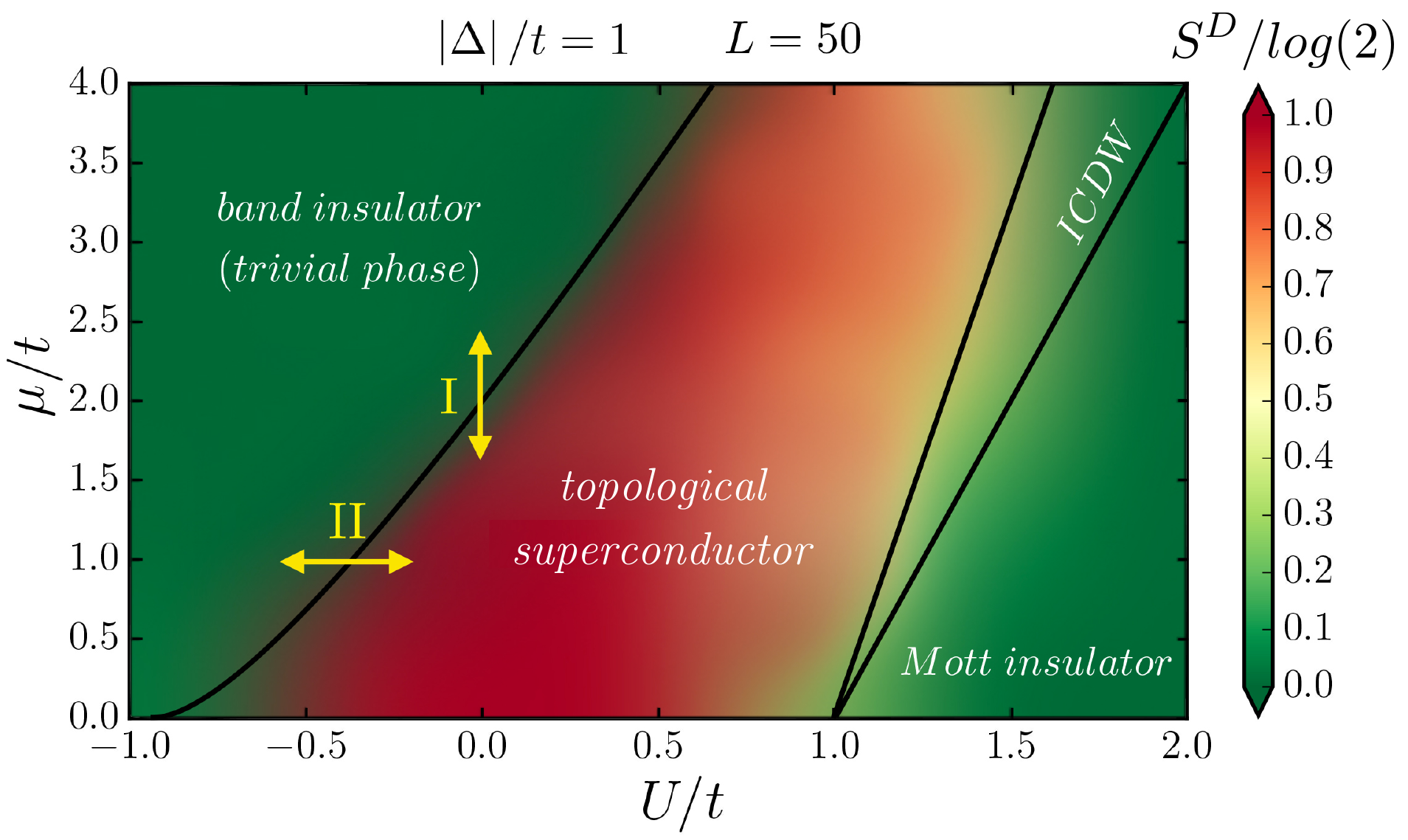}
\put (0,50) {b)}
\end{overpic}
  \caption{ \label{fig:cuts}(Color online) Partitions associated with the entanglement topological order parameter and the phase diagram of the interacting Kitaev chain. Panel \textit{a)}: schematics of the partitions $A$ (shaded, green) and $B$ (shaded, blue) considered here. Each site of the chain hosts a spin-less fermion degree of freedom $a_j$, that can be decomposed into two Majorana fermions $c_{2j}$ and $c_{2j+1}$. The orange circle magnifies the cut across partitions: deep in the topological phase, neighboring Majorana fermions belonging to different physical sites are coupled (dashed line). The partition cut takes place exactly between the two coupled Majorana fermions. Panel \textit{b)}: disconnected von Neumann entropy $S^D$ as a function of $\mu/t, U/t$, at fixed $\Delta=1$. Black lines are from Ref.~\citen{Katsura15}. The color plot is obtained via interpolation on a 5 x 7 grid. Clearer evidence of the quantization of $S^D$ in a phase and the sharpness of the transition requires a lot of points, as we highlight on the lines I and II in Sec.~\ref{sec:univ}.
}
\end{figure}

In this work, we show how entanglement and R\'enyi entropies of \textit{disconnected partitions} provide a set of entanglement order parameters for one-dimensional topological superconductors (TSCs)~\cite{Kitaev01,Beenakker:2013aa}. These order parameters satisfy the following properties: \textit{(i)} they are quantized to 0 or $\log 2$ when the phase is topologically trivial or not-trivial respectively, and are thus able to detect the single entanglement bit - an ebit - that can be distilled from the ground state manifold; \textit{(ii)} they display a scaling behavior when approaching quantum phase transitions, thus defining entanglement critical exponents that describe the build-up of non-local quantum correlations across such transitions; \textit{(iii)} some of them are experimentally measurable in- and out-of-equilibrium utilizing recently introduced~\cite{Elben_2018,Vermersch_2018} and demonstrated~\cite{Brydges:2019aa} techniques based on random measurement methods~\cite{Enk:2012aa}. 

Following Ref.~\citen{Casini:2004ab}, we consider the $F$-function between two partitions $A$, $B$, which compensate for all edges and volume contributions in an open chain of length $L$. These properties are required to avoid non-universal effects: the significance of our diagnostics relies on an underlying gauge theory description below, which calls for quantities that are divergence-free in the continuum limit. To diagnose the presence of non-local correlations in the system, we choose two partitions with {\it different connectivity}, as shown in Fig.~\ref{fig:cuts} a). 
The resulting disconnected $n$-entropies $S^D_n$ read:
\begin{equation}
S^D_{n}= S_{A, n} + S_{B, n} - S_{A \cup B, n} - S_{A \cap B, n},
\label{eq:sqtopo}
\end{equation}
where $S_{F, n}$ is the bipartite R\'enyi entropy of order $n$ of the partition $F=A,B,A \cup B$ or $A \cap B$. We define $L_{\alpha}$ as the size of a given partition and $L_D$ as the distance between the two different parts of $B$. Unless stated otherwise, we consider the representative case $L_A=L_B$, which provides a cleaner finite-size scaling analysis.

We denote the case $n=1$ as $S^D$, that corresponds to the von Neumann entropy, and satisfies $S^D>0$ because of strong subadditivity. This entropy improves from Ref.~\citen{Wang2015} that uses two systems with different boundary conditions. It has been considered in Ref.~\citen{Zeng:aa}, which pointed out a strong analogy between bosonic symmetry-protected topological phases (SPTPs) and error-correcting codes. Here, we focus instead on fermionic phases where topology stems from an underlying fundamental symmetry (parity) which cannot be broken by any Hamiltonian perturbation. This condition plays a crucial role in defining the upcoming gauge-theory picture describing the entanglement content of such states while the non-local correlations introduced by the fermionic algebra via the Jordan-Wigner string are responsible for such analogy. 

At a qualitative level, the key in $S^D_{n}$ is the disconnected partition $B$: all other terms are complementary, and only required to eliminate non-universal boundary and volume terms. In Fig.~\ref{fig:cuts} b), we show the finite-size behavior of $S^D$ across the phase diagram of interacting Kitaev chains: this plot illustrates graphically how, even at modest system and partition sizes, $S^D$ clearly distinguish topological from trivial phases. We note that Ref.~\citen{Kim2014} computes a quantity coinciding with $S^D_n$ for the non-interacting Kitaev chain, but it should deviate when a quantum spontaneous symmetry breaking phase is involved.

The present study is divided into the theoretical part Sec.~\ref{sec:theory} and the computational part Sec.~\ref{sec:numeric}. The first part introduces shortly the Kitaev model (Sec.~\ref{sec:shortmodel}). Then, it both derives and gives an intuitive picture of entanglement and the topological order parameter $S^D$ in the deep topological or trivial regime (Sec.~\ref{sec:extreg}) and using lattice gauge theories (Sec.~\ref{sec:gauge}). The computational part displays the efficiency of $S^D$ as a topological detector (Sec.~\ref{sec:phasediag}), reveals its universal behavior at the transition (Sec.~\ref{sec:univ}), shows its response to quenches which is typical of a topological invariant (Sec.~\ref{sec:topoinvar}), and confirms its robustness to symmetry-preserving disorder (Sec.~\ref{sec:disorder}). We then discuss the existing experimental relevancy of the detector in Sec.~\ref{sec:exp}, before concluding. The appendices mirror the main structure of the article and add miscellaneous details referred to in the text. In particular, appendix~\ref{sec:Kitaevplus} provides a longer introduction on the Kitaev wire for the unfamiliar reader while appendix~\ref{sec:equiv} attempts a proof of the equivalence between all $S^D_{n}$ as topological detecting quantities.

\section{Disconnected entropy of the interacting Kitaev wire}\label{sec:theory}

The Kitaev wire gives a prime example of a topological phase in 1D, with well-known behavior deep into either the topological or trivial phase. There, a topological signature is more easily extracted and identified from the entanglement of a ground state. An analytical treatment is possible and presented here.

\subsection{The Model Hamiltonian.} \label{sec:shortmodel}
We consider the interacting version of Kitaev p-wave superconductor, whose Hamiltonian reads:
\begin{equation}
\begin{split}
H&=\sum_{j=1}^{L-1}\left[-t (a_j^\dagger a_{j+1}+\textrm{h.c.})+ (\Delta a_j a_{j+1}+\textrm{h.c.}) \right. \\
&+\left.4U\left(n_j-\frac{1}{2} \right)\left(n_{j+1}-\frac{1}{2} \right)\right]-\mu\sum_{j=1}^{L}n_j,
\end{split}
\label{eq:hamKitint}
\end{equation}
where $a_j^\dagger$ ($a_j$) are the creation (annihilation) operators of the spinless fermion on site $j$, $n_j=a^\dagger_ja_j$, $t$ is the hopping amplitude, $\Delta$ is the superconducting amplitude, $U$ is the nearest-neighbor interaction, and $\mu$ is the chemical potential. The phase diagram of the model is known~\cite{Stoudenmire11, Katsura15} and displays a TSC phase, in addition to topologically trivial phases, including a band insulator, a Mott insulator, and an incommensurate charge-density-wave (ICDW) phase. For any state in the Hilbert space, the bipartite properties of a simply connected partition are equivalent to the ones of the XYZ spin chain obtained from Eq.~\ref{eq:hamKitint} after applying a Jordan-Wigner transformation. As such, they are uninformative about the topological origin of a given phase. 

\subsection{Disconnected entropies at exactly soluble points} \label{sec:extreg}
Therefore, the goal is to find a combination of entropies able to both unambiguously capture the influence of non-locality in the ground state properties and identify the amount of information - in this case, a single ebit - that can be stored in the ground state manifold. The goal is reached for $S^D$ which contains the simplest non-trivial disconnected partition, $S_{B}$ while all other terms  only compensate possible volume and edge effects. 

For conformal phases, $S^D_n$ is immediately given by conformal field theory~\cite{Caraglio:2008aa,Furukawa:2009aa,Calabrese_2009}, and vanishes in the thermodynamic limit. For gapped phases, one has to distinguish between topologically trivial and non-trivial phases. We analyze here the limiting cases.

\textit{(i) $t=\Delta=U=0, \mu>0$}: the system is a band insulator, and the density matrix of arbitrary partitions has rank 1 both in fermionic and spin systems. This immediately gives $S^D_n=0$. The same result holds in the Mott insulator phase. Thus, the thermodynamic limit of $S^D_n$ cannot distinguish conformal phases and trivial phases. Only the finite size correction of $S^D_n$ may distinguish them.

\textit{(ii) $t=\Delta=1\gg|\mu|, U=0$}: this regime is representative of the TSC phase. Its correspondent in the XYZ model is a ferromagnetic phase, which we analyze first as a representative of a symmetry-broken phase. There, the lowest energy states at any finite size are equal weight superpositions of the two ferromagnetic states, $|\Psi_{XYZ}\rangle = (|\uparrow\uparrow\uparrow...\rangle\pm|\downarrow\downarrow\downarrow...\rangle/\sqrt{2}$, separated by a gap $\delta\propto e^{-L}$. For both states, any reduced density matrix of an arbitrary spatial partition is equivalent, and thus $S^D_n=0$. 

For the TSC, the situation is different. While $S_A, S_{A\cap B}, S_{A\cup B}$ are the same as in the spin model, $S_{B}$ has a sharply different behavior. In this regime, the ground state is two-fold degenerate (again, up to a gap $\delta\propto e^{-L}$): Each of the two states $|\Psi\rangle_\pm$ can be written as an equal weight superposition of states with a given parity $|\psi\rangle_\pm$, i.e., $|\Psi\rangle_\pm = (1/2^{L-1})\sum_\psi|\psi\rangle_\pm$. The proper fermionic trace evaluates the entanglement structure of arbitrary partitions so that we obtain $S_{B}=2\log 2$. This returns a disconnected entropy $S^D = \log 2$. A full derivation is given in appendix~\ref{sec:entprop}.

The behavior of these cartoon wave functions sharply distinguishes the TSC phase with respect to all other phases. We need to go beyond these cartoons to understand if this behavior is a property of a phase, and if the value of $S^D$ remains quantized in the whole TSC phase. Before presenting numerical results in support of these findings, we now illustrate how the quantization of $S^D$ emerges naturally when utilizing a lattice gauge theory (LGT) description of the Kitaev chain.

\subsection{Gauge theory characterization of entanglement properties.} \label{sec:gauge}
\begin{figure}[t]
   \includegraphics[width=.48\textwidth]{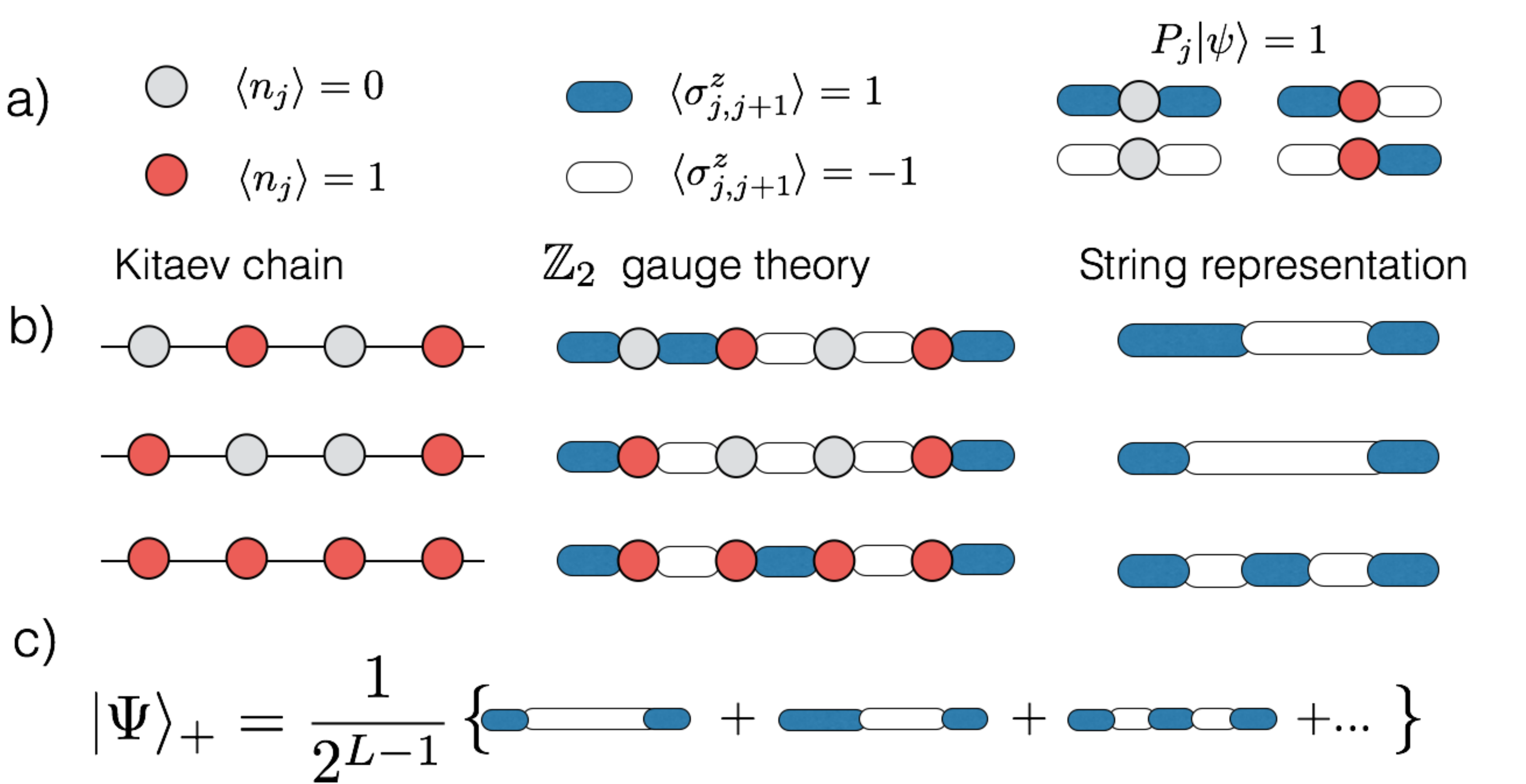}
  %\end{minipage}
  \caption{ \label{fig:Z2LGT}(Color online) Schematics of the correspondence between the Kitaev chain and $\mathbb{Z}_2$ lattice gauge theories. \textit{a)}: Hilbert space structure and gauge-invariant building blocks. \textit{b)}: three examples of the mapping between states in the fermionic (left), gauge theory (center), and string representation (right). \textit{c)}: string representation of $|\Psi\rangle_+$.
}
\end{figure}

The starting point is the exact relation~\cite{MCCOY1983278} between Eq.~\ref{eq:hamKitint} and a $\mathbb{Z}_2$ LGT, that we schematically review. The  $\mathbb{Z}_2$ LGT describes the coupling between the $\mathbb{Z}_2$ gauge fields residing on bonds (represented here by Pauli matrices, $\sigma^\alpha_{j,j+1}$), and the hard-core Higgs fields $\varphi_j$, with $n_j=\varphi_j^\dagger\varphi_j$, defined on the vertices. The gauge-invariant Hilbert space is defined as the set of states where the local parity $P_j = (1-2n_j ) \sigma^z_{j-1,j}\sigma_{j,j+1}^z$ is fixed to 1 (see Fig.~\ref{fig:Z2LGT}). Under open boundary conditions (OBC), we fix to $\sigma^z_{0,1}=1$ the value of the first gauge field without loss of generality. The value of the last gauge field $\sigma^z_{L,L+1}=P$ gives the total parity of the system due to gauge invariance.

The ground state wave functions $|\Psi\rangle_\pm$ can be described in terms of either fermionic or gauge fields, since, in 1D, those are mutually fixed by Gauss law. In terms of LGT, the ground states are equal-weight superpositions of all possible string states of arbitrary length, and compatible with the boundary conditions: a sample of those are depicted in Fig.~\ref{fig:Z2LGT} for the case $P=1$. This picture describes a 1D gauge theory in a phase with strongly fluctuating gauge fields, and is strongly reminiscent of the loop description of 2D quenched $\mathbb{Z}_2$ LGT~\cite{Lacroix2010,Zeng19,Wen19}. 

Evaluating entanglement entropies in this phase is straightforward by exploiting gauge invariance: 

\textit{i)} The entropy of each connected partition is $\log 2$. Indeed, let us define $\sigma_L^z, \sigma^z_R$ as the two boundary spins of the partition. Their product is equal to the parity of the partition: the density matrix of the partition is block-diagonal in this conserved quantum number. If the correlation length is much smaller than the partition length, both positive and negative parities are equally probable and all states count with equal weight. The corresponding von Neumann entropy is thus $\log 2$.

\textit{ii)} the entropy of disconnected partitions is $(N_c-1)\log 2$, where $N_c$ is the number of partitions such that $N_c=2$ corresponds to the standard bipartition of the chain into two halves. Indeed, let us define as $\sigma_{L,h}^z, \sigma^z_{R,h}$ the gauge fields at the boundaries of the partition $h$. As long as the length of each partition is larger than the correlation length, each patch is an equal-weight superposition of all possible states, under the condition that $P_h=+1$ or $-1$ for partitions with or without an out-coming flux respectively. For a fixed total parity, there are $2^{N_c-1}$ finite, equal values of the corresponding density matrix, which returns an entropy equal to $(N_c-1)\log 2$.

We emphasize that the gauge theory description enables a simple calculation of the entropies (by replacing fermionic statistics with a $\mathbb{Z}_2$ gauge field), and, at the same time provides a simple, compelling physical picture, that might be extended to more exotic types of order.

\section{$S^D$ as a topological invariant order parameter} \label{sec:numeric}

$S^D$ is efficient as a topological order parameter even for modest system size. Indeed, we simulate the quantity and obtain a sharp phase diagram of the Kitaev wire. Sharp, because close to the phase transition, it displays universal behavior typical of an order parameter and allows definition of critical entanglement exponents. Its characteristics as a topological invariant are confirmed by its invariance after a quench as expected from Sec.~\ref{sec:gauge} and by its robustness to disorder.

\subsection{Scaling of $S^D$ and phase diagram} \label{sec:phasediag}
We therefore turn to the numerical investigation of Eq.~\ref{eq:hamKitint}. We used free fermion techniques~\cite{Peshel03} to investigate the non-interacting case $U=0$, and density-matrix-renormalization group (DMRG)~\cite{White1992,Schollwock2005} for $U\neq0$. Since DMRG does not give immediate access to $S_{B}$, we performed separate simulations to obtain this quantity, by modifying the lattice connectivity (cf appendix~\ref{sec:DMRG}). We kept at least 1200 states after truncation and performed at least 30 sweeps. Typical discarded weights at the end of the simulation were of order $10^{-8}$.

\begin{figure}[]
\centering
\begin{overpic}
[width=.9\linewidth]{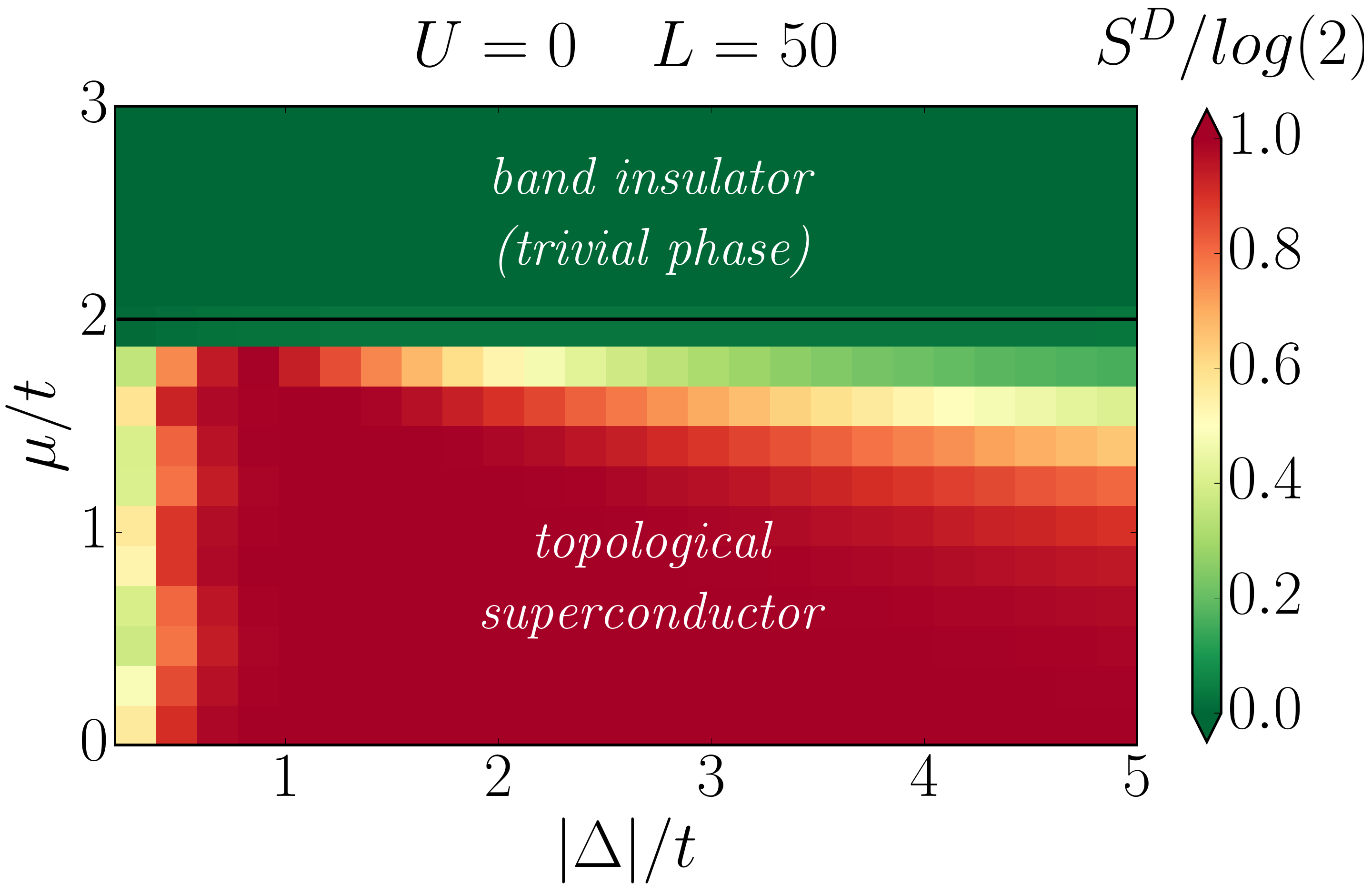} \ \  \
\put (10,60) {(a)} 
\end{overpic}\hfill
\begin{overpic}
[width=.9\linewidth]{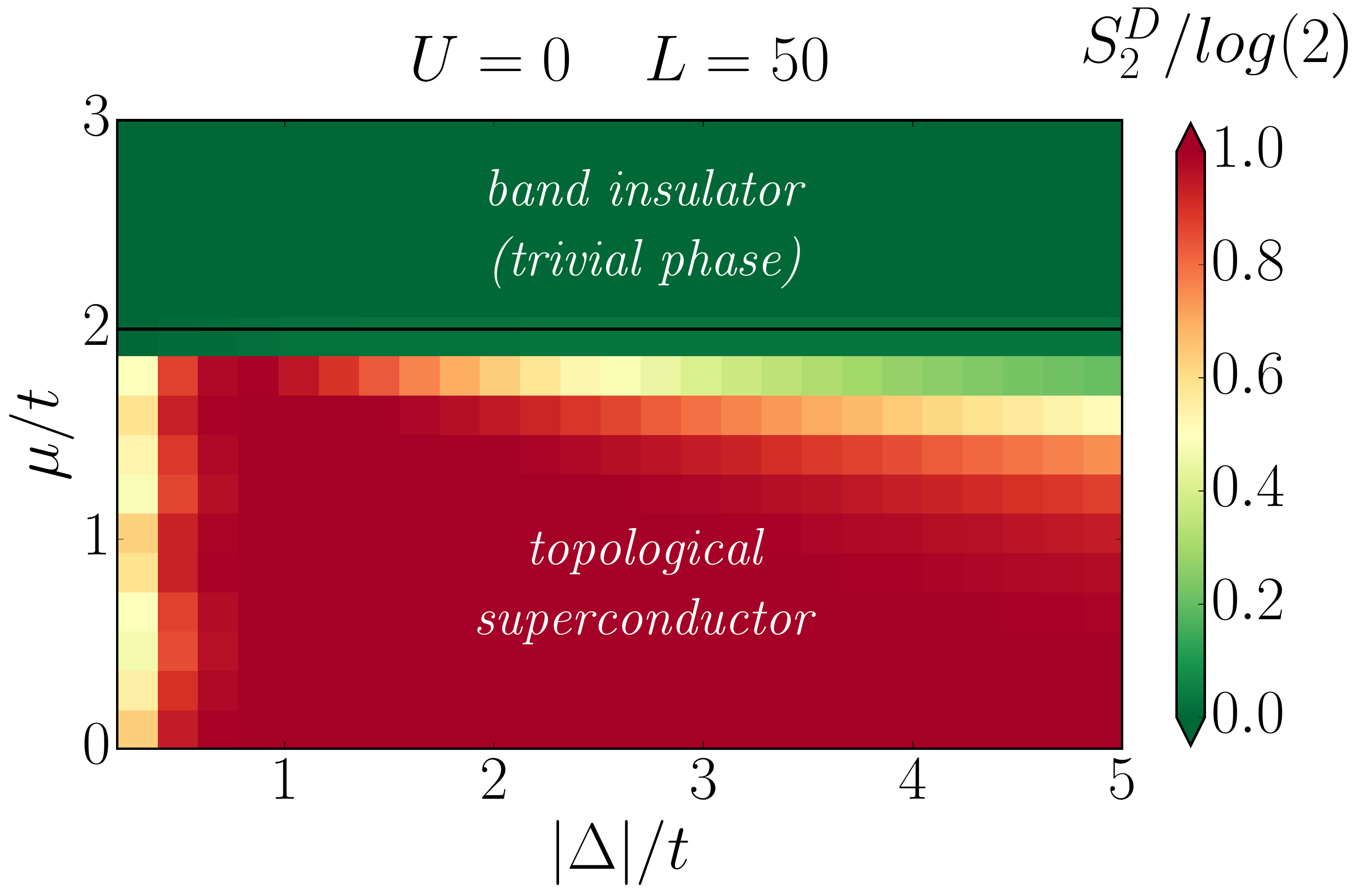}
\put (10,60) {(b)} 
\end{overpic}
\caption{Color plot of the disconnected n-entropies, $S_n^{D}$, of the Kitaev wire without interaction, $U=0$, for (a) $n = 1$ and (b) $n = 2$. 
The results are obtained with the free-fermion technique.
The $y$-axis represents the chemical potential,  $\mu/t$, while the $x$-axis the superconducting amplitute, $\lvert \Delta \rvert/ t$.  
A grid with $16 \times 30$ points is considered.
The two theoretically expected phase transition lines occur for $\lvert \Delta \rvert = 0$ and $\mu / t = 2$, respectively.
Using the definitions of Fig.~\ref{fig:cuts} a), $L = 50$, $L_A =L_B = 12$ and $L_D = 32$. The diagrams coincide with the results obtained in Ref.~\citen{Kim2014}.}
\label{fig:KitaevU0}
\end{figure}

\begin{figure}[t]
   \begin{overpic}[width=1.\linewidth]{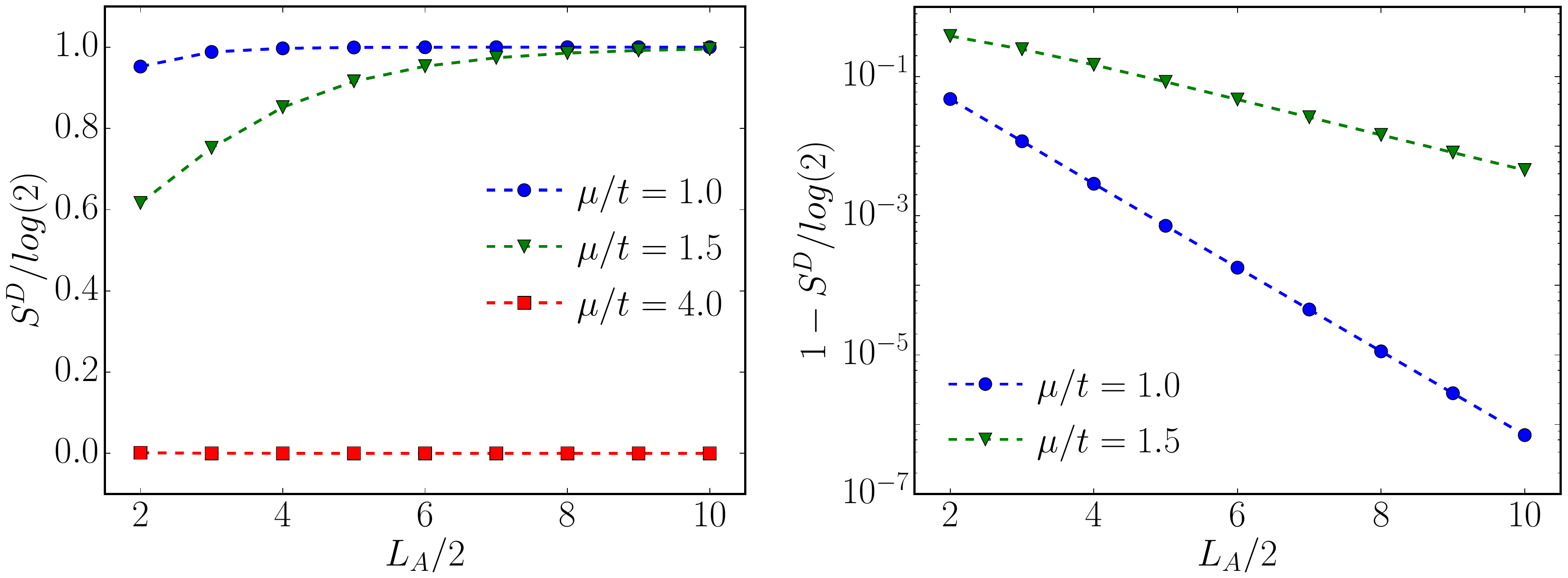}
   \put (0,31) {a)} \put (50,31) {b)}
\end{overpic}\hfill
  \caption{ \label{fig:FSS} (Color online) Finite-size scaling properties of $S^D$ in a) normal scale, b) log scale, for a chain with $L_A=L_B=(L_{A \cap B}+L_{A \cup B})/2$, and $U=0, \Delta/t=1$. In the topologically trivial phase, $S^D$ quickly vanishes. In contrast, in the topological phase ($\mu=1.0, 1.5$), $S^D$ increases as a function of system size, and approaches its thermodynamic value exponentially fast when increasing $L_A$, as shown in b).}
\end{figure}

The phase diagrams in Figs.~\ref{fig:cuts} b) and~\ref{fig:KitaevU0} a) (with and without interactions) show how, even at very modest partition sizes, $S^D$ is large and finite only the TSC phase. Comparison between Figs.~\ref{fig:KitaevU0} a) and b) shows the equivalence between $S^D$ and $S^D_2$. In Fig.~\ref{fig:FSS} a), we show the finite-size-scaling behavior of $S^D$ for representative points in the TSC ($\mu/t=1.0, 1.5$) and topologically trivial ($\mu/t=4$) phase. The asymptotic values are quantized within numerical accuracy of our fits to $\log 2$ and 0, respectively, in agreement with the theoretical discussion above. Fig.~\ref{fig:FSS} b) shows how, in the TSC phase, quantization is approached exponentially fast in system size; the same holds true for $S^D_n$. 
%%%%%%%%%%%%%%%%%%%%%%%%%%%%%%%%%%%%%%%%%%%%%%%%%%%%%%%%
%%%%%%%%%%%%%%%%%%%%%%%%%%%%%%%%%%%%%%%%%%%%%%%%%%%%%%%%

\begin{figure}[t]
  \centering
   \begin{overpic}[width=1.\linewidth]{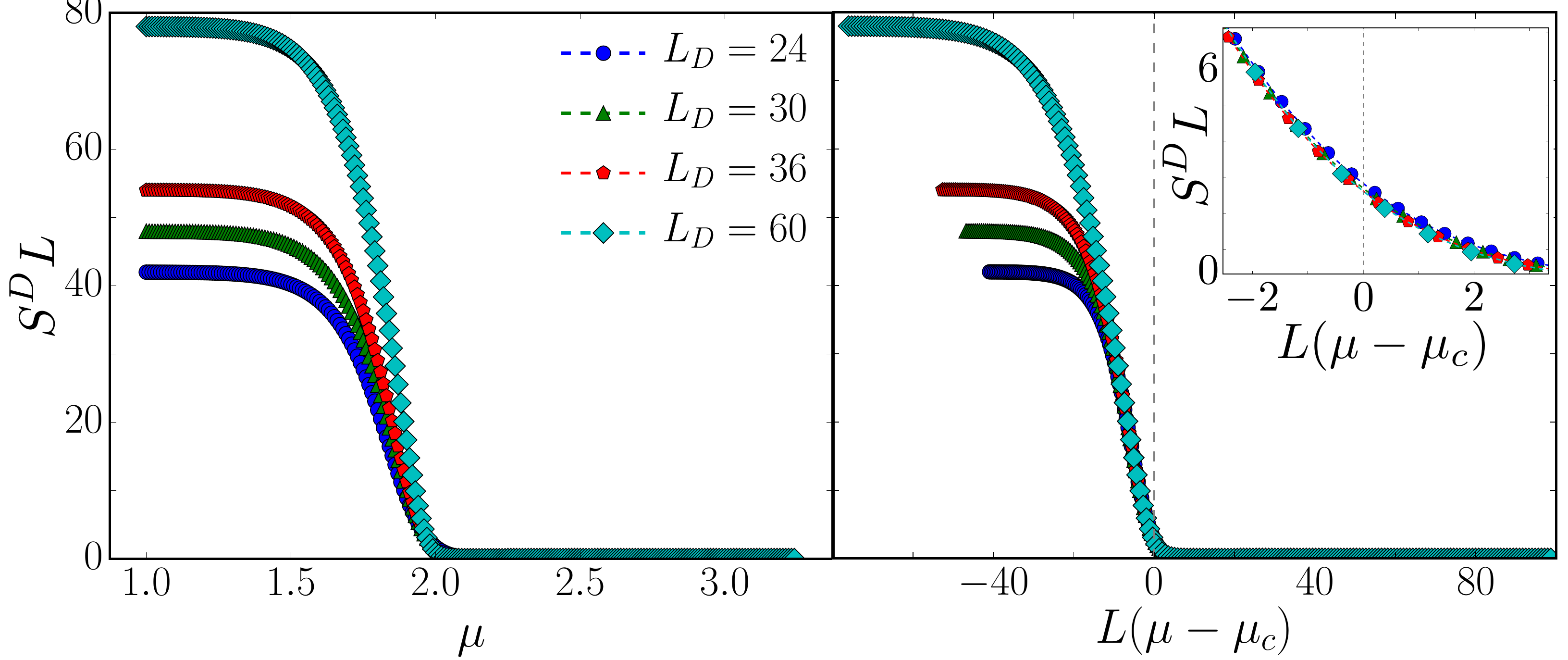}
 \put (10,35) {a)} \put (55,35) {b)}
\end{overpic}\hfill
    \begin{overpic}[width=1.\linewidth]{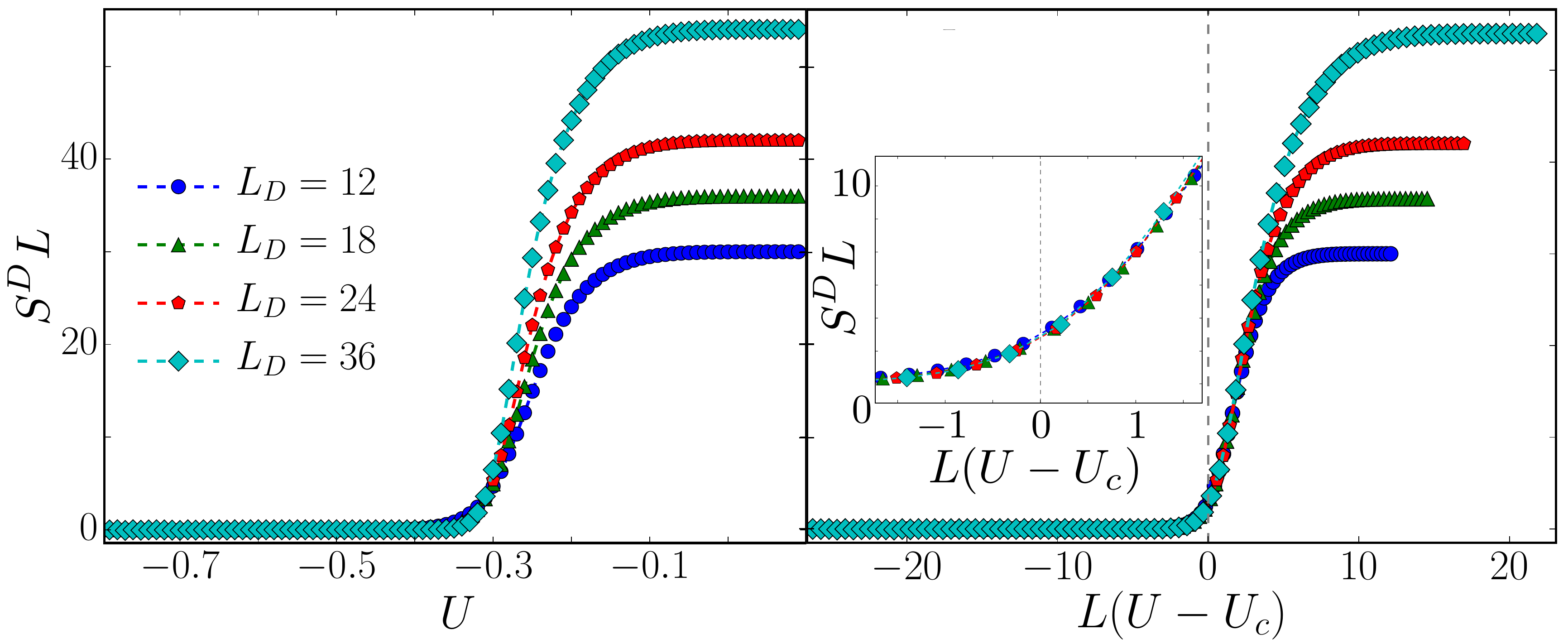}
    \put (10,35) {c)} \put (55,35) {d)}
\end{overpic}
  \caption{ \label{fig:scaling} Finite size scaling of $S^{D}$ (in units of $\log_2$) along the line (I) and (II) of Fig.~\ref{fig:cuts} as a function of $L_D$ and $\mu$ or $U$ using DMRG. In all plots, $L_A = L_B = 12$. (a) $S^{D}L$ as a function of $\mu$ for different sizes: the critical value $\mu_c$  is the intersection of all curves; we obtain $\mu_c = 1.978$. (b) Scaling of $\lambda(x)$ for different system sizes: curve collapse. The collapse is best realized for $a = b = 1$, values that also minimize the square root of the residual sum of squares.(c) $S^{D}L$ as a function of $U$ for different sizes: the critical value is here $U_c(12)=-0.314$. (d) The collapse is again best realized for $a=b=1$. Simulations with more sites (especially using free fermion techniques) only confirm these results.}
\end{figure} 

\subsection{Universal behavior and entanglement critical exponents.}  \label{sec:univ}
Since $S^D$ captures universal properties of each phase, it is natural to wonder whether such quantities can display universal scaling behavior when crossing a quantum phase transition.  Here, we focus on the transition between TSC and band insulator, which belongs to the Ising universality class. 

Similarly to conventional quantum critical behavior, we fit $S^D$ using a phenomenological finite-size scaling ansatz:
\begin{equation}
S^{D}L^{\frac{a}{b}}=\lambda\left(L^{\frac{1}{b}}(\alpha-\alpha_{c})\right),\label{eq:scaling_formula}
\end{equation}
where $\alpha= \mu$ or $U$ is the varying parameter chosen, $a$ and $b$ (\textit{a priori} different depending on the chosen parameter) take place of the usual critical exponent $\beta$ and $\nu$, and $\lambda(x)$ is a scaling function. We extract these parameters using curve intersections and collapse shown in Figs.~\ref{fig:scaling} across the transition indicated in Fig.~\ref{fig:cuts} with the yellow arrows (I) and (II). The results of the collapse scaling locates correctly the transition point (with errors ~$10^{-4}$). Most surprisingly, we find that the entanglement critical exponents satisfy $a = b =1$ irrespectively of where the transition line is crossed, a sharp signature of universal behavior. The quality of the collapse scaling in the interacting case is already good for modest system sizes, further corroborating such universal behavior.

\subsection{Invariance of $S^D$ under coherent dynamics.} \label{sec:topoinvar}
In the thermodynamic limit, topological invariants cannot change under unitary evolution (as long as specific symmetries are not broken explicitly~\cite{Caio:2015aa,DAlessio:2015aa,McGinley:aa}). To check that $S^{D}$ is a true topological invariant, we performed an extensive investigation based on quantum quenches within and across the topological phase. 

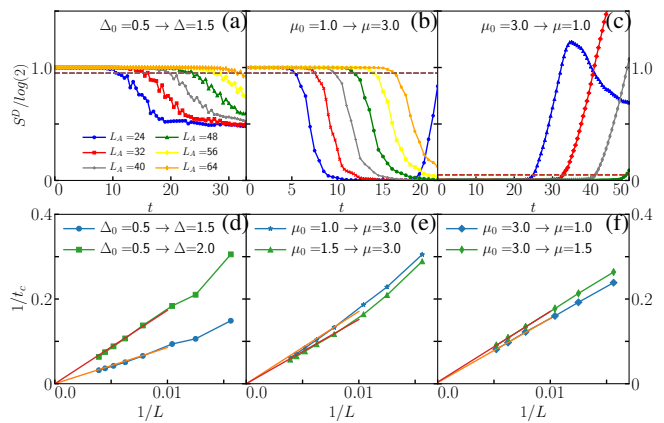
\begin{figure}
\centering
   \begin{overpic}[width=1.\linewidth]{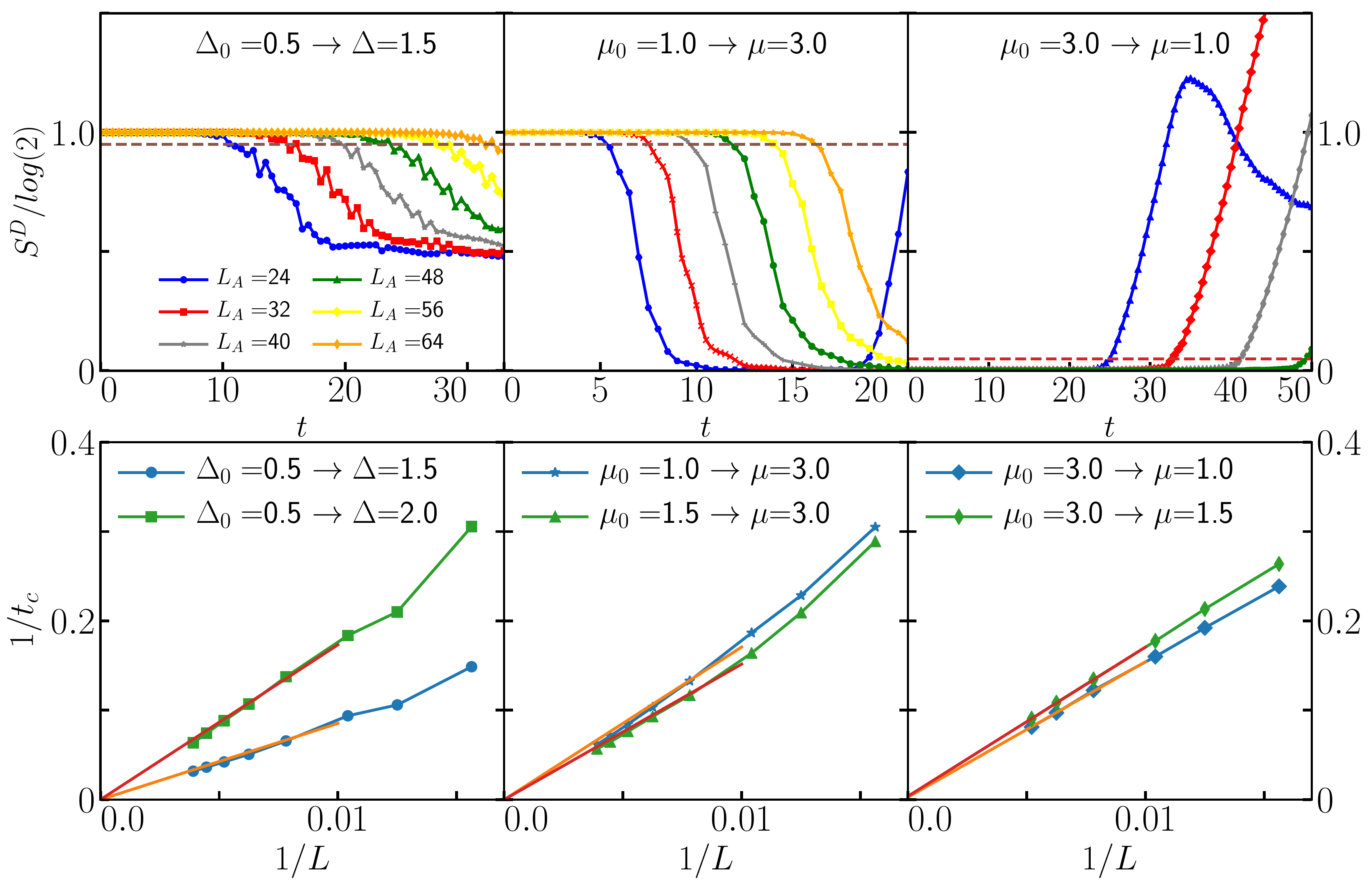}
 \put (33,61.2) {(a)} \put (62,61.2) {(b)} \put (92,61.2) {(c)}
 \put (33,30.2) {(d)} \put (62,30.2) {(e)} \put (92,30.2) {(f)}
\end{overpic}
  \caption{ \label{fig:quenches} (Color online) Time evolution of $S^D$ after a quantum quench from a) $\Delta_0=0.5$ to $\Delta=1.5$, b) from $\mu_0=1.0$ to $\mu=3.0$ and c) from $\mu_0=3.0$ to $\mu=1.0$ with $U=0,\mu=0,L=8L_A$. Finite-size scaling of $t_c$ or $t_c^\prime$ for two values of d) $\Delta$, e) $\mu_0$ and f) $\mu$ of the quenched Hamiltonian. In all cases, the width of the plateau diverges linearly with system size, as expected for topological invariants. The threshold lines of $S^D$ according to the definition of $t_c$ are depicted as a dashed line.}
\end{figure}

A representative sample of our results is presented in Figs.~\ref{fig:quenches}. In panel \textit{a)}, we plot the time evolution of $S^{D}$ for a quench from an initial value of the superconducting amplitude $\Delta=0.5$ to a final value $\Delta=1.5$. Different lines correspond to different system sizes. For each size, one can sharply distinguish two regimes. At short times, $S^{D}$ does not change with time and exhibits a plateau up to a time $t_c$ that depends on $L_A$. After this timescale, quantization is lost, and the dynamics is dictated by non-universal dynamics. To understand whether quantization is a robust feature, we perform a finite-size scaling analysis in panel \textit{d)}: our results show that $t_c$ (defined as the time when $S^D=0.95$) grows approximately linearly with system size and will diverge at the thermodynamic limit. This behavior confirms the topological invariant nature of $S^D$. Panels \textit{b)}, \textit{e)}, and \textit{c)}, \textit{f)} confirms these results for the quench $\mu_0=1.0$ to $\mu=3.0$ (from topological to trivial) and $\mu_0=3.0$ to $\mu=1.0$ (from trivial to topological). For f), $t_c$ is the time when $S^D=0.05$.

\subsection{Robustness of $S^D$ to disorder}\label{sec:disorder}

In 1D, the effect of disorder is the most drastic: localization of the wave function occurs as soon as disorder exists, and not even diffusive transport is possible~\cite{Giamarchi2004}. Topological insulators and superconductors escape the effect, in the sense that their extended edge states stay robust against symmetry-preserving disorder. The same disorder should also preserve $S^D$ that relies on these edge states. Numerical simulations indeed confirm the robustness of $S^D$ to disorder for the non-interacting Kitaev wire. 

\begin{figure}[t]
  \centering
   \begin{overpic}[width=.9\linewidth]{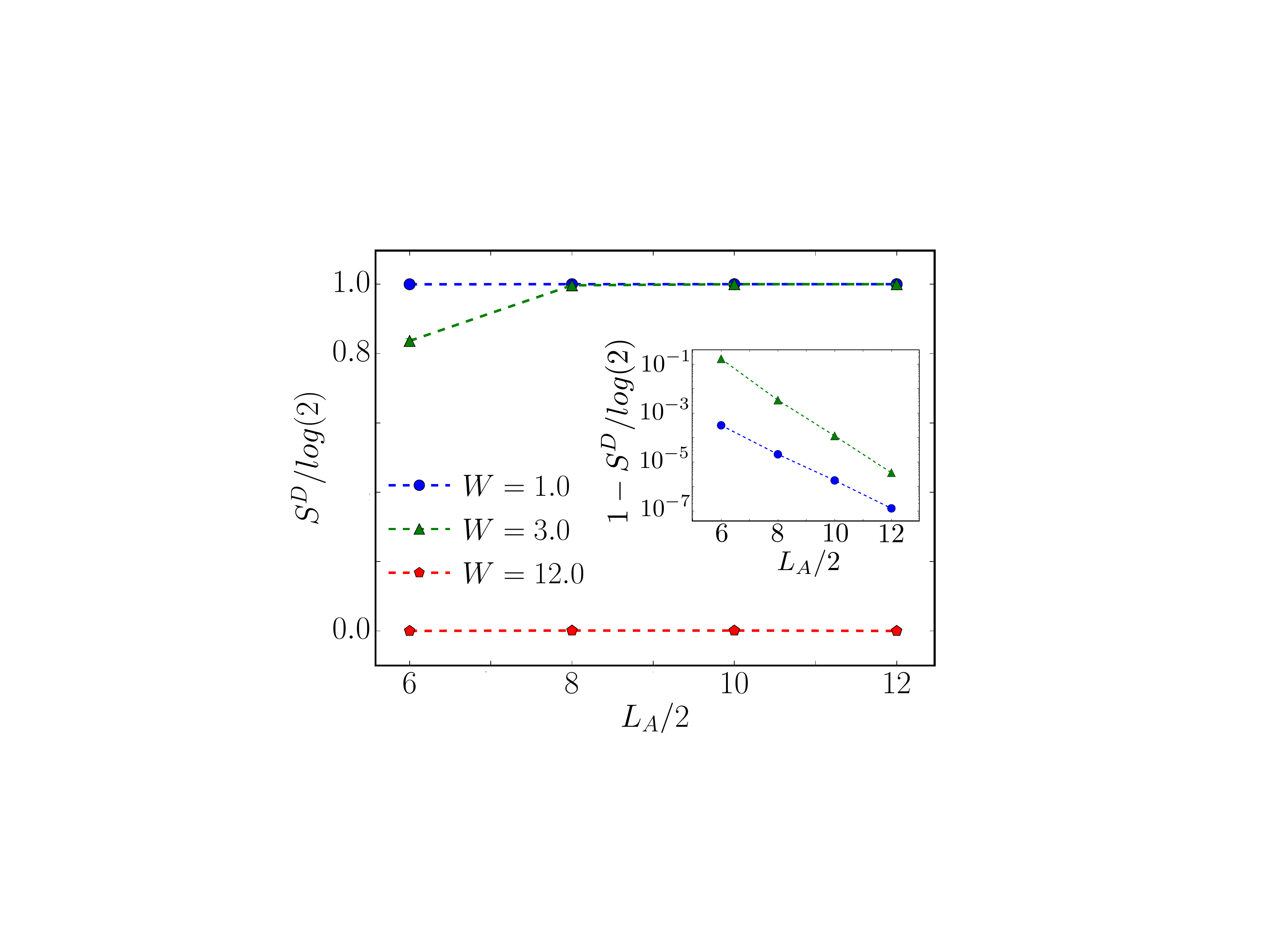}
   \end{overpic}
    \caption{Scaling in system size of the mean value of SD for $t=\Delta=\mu=1$, $U=0$, $L_A=L_B=(L_{A \cap B} + L_{A \cup  B} )/ 2$ and for three amplitudes of disorder: $W=1$ and 3 in the topological phase, and $W=12$ for the trivial disordered phase (200 realizations of the disorder for each point). The inset provides a logarithmic scale for the axis of $S^D$: the scaling is exponential. The standard deviation is smaller than $10^{-6}$ for each point. The same study using R\'enyi-2 entanglement entropies leads to the same results quantitatively. The phase transition is expected at $W_c \sim 11$ for these parameters~\cite{Gergs2016,Levy2019}.}
    \label{fig:disorder}
\end{figure}

We introduce finite Anderson-like disorder using:
\begin{equation}
\mu_i= \mu + \delta_i, \;\;\;\; \delta_i \in \left[ -W/2 ; W/2 \right],
\end{equation}  
where $\mu_i$ is the new position-dependent ($i$) chemical potential in the Hamiltonian Eq.~\ref{eq:hamKitint} (with $U=0$). $\delta_i$ is a random variable of uniform distribution, and $W$ is the amplitude of the disorder. The new potential breaks the translation symmetry, but not the protecting symmetries of the topological phase that persists for a reasonable amplitude of the disorder. In Fig.~\ref{fig:disorder}, we draw the mean value of $S^D$ over realizations of disorder for different $W$ as we increase the system size $L$. $S^D$ scales exponentially in system size towards the quantized value of $\log 2$ for the topological phase, 0 otherwise.

\section{Discussion \& Conclusions.}

\subsection{Experimental measurement and comparison to other diagnostics.} \label{sec:exp}
The probes $S^D_n$ are experimentally-relevant because they are already informative for modest partition sizes, and because R\'enyi entropies can be measured. The proposals in Ref.~\citen{Elben_2018} discuss how to perform measurements of R\'enyi-2 entropies in synthetic quantum systems: the complexity of the measurements is not sensitive to the connectivity of the partition itself, but only to its total size. Given that a large $L_D$ allows the distillation of the correct information from the wave function, measuring $S^D$ is as complex as measuring its largest partition $A$. We note that partitions of sizes up to 10 spins have already been probed in experiments~\cite{Brydges:2019aa}. 

Finally, we comment on the relation between $S^D$ and other diagnostics. Topological invariants such as the many-body Chern number~\cite{Niu1985,fradkinbook} are unrelated to bipartite entanglement properties, as they do not depend solely on the spectrum of density matrices, but also on their eigenfunctions. For specific symmetries, specific topological invariants can be defined~\cite{Chen:2012aa,Haegeman:2012aa,Pollmann:2012aa,Shapourian2017b,Shapourian2017a,Zeng19} (and potentially experimentally measured~\cite{Elben:aa}) also utilizing the matrix-product-state (MPS) classification of SPTPs; these quantities are genuinely sensitive to the response of a state to specific (symmetric) operations, and not immediately connected to entanglement. From a theoretical viewpoint, all these diagnostics represent complementary tools, that give access to qualitatively different features characterizing topological matter. Examples now include: response of wave functions under changing boundary conditions (Chern number), properties with respect to protecting symmetry (MPS order parameters), and non-local entanglement content of wave functions (disconnected entropies).

\subsection{Conclusions.} We have shown how the entanglement of disconnected partitions uniquely distinguishes topological superconducting phases in one-dimensional systems. 
This distinction is naturally interpreted within a lattice gauge theory framework, and leads to key footprints both at the ground state level, and in quantum quenches. Entanglement order parameters display universal scaling behavior when crossing phase transitions, characterized by entanglement critical exponents. 
Our findings show that modest partition sizes - of the order of what has been already experimentally demonstrated - are sufficient to uniquely characterize topological superconductors via entanglement. 
It would be intriguing to investigate whether other forms of quantum correlations between disconnected partitions, such as discord~\cite{Ollivier:2001aa} or quantum coherences~\cite{Frerot:2016aa}, display similar characteristic features, and if entanglement topological invariants can be used to characterize the real-time dynamics of interesting topological matter~~\cite{McGinley:aa}.

%%%%%%%%%%%%%%%%%%%%%%%%%%%%%%%%%%%%%%%%%%%%%%%%%%%
\begin{acknowledgments}
%%%%%%%%%%%%%%%%%%%%%%%%%%%%%%%%%%%%%%%%%%%%%%%%%%%
We thank S. Barbarino, G. Giudici, F. Illuminati, N. Lindner, L. Pastori, D. Schuricht, F. Surace, X. Turkeshi, and B. Vermersch for useful discussion, and J. Budich for suggesting the investigation of the real-time dynamics. This work is partly supported by the ERC under grant number 758329 (AGEnTh), and has received funding from the European Union's Horizon 2020 research and innovation programme under grant agreement No 817482. G.M. is partially supported through the project ``QUANTUM'' by the Istituto Nazionale di Fisica Nucleare (INFN) and through the project ``ALMAIDEA'' by the University of Bologna.
\end{acknowledgments}

%%%%%%%%%%%%%%%%%%%%%%%%%%%%%%%%%%%%%%%%%%%%%%%%%%%%
\appendix
%%%%%%%%%%%%%%%%%%%%%%%%%%%%%%%%%%%%%%%%%%%%%%%%%%%%

\section{Additional information on the Kitaev model}

In this section, we briefly present the Kitaev model and detail the derivation of all the analytical results mentioned in the main text. More specifically, we focus on the regime described in Kitaev's original paper~\cite{Kitaev01} whose algebra is simpler while containing important features on the entanglement properties of the whole model when it displays a topological phase.

\subsection{The Kitaev model without interaction}\label{sec:Kitaevplus}

\begin{figure}[htb]
   \includegraphics[width=1.0\linewidth]{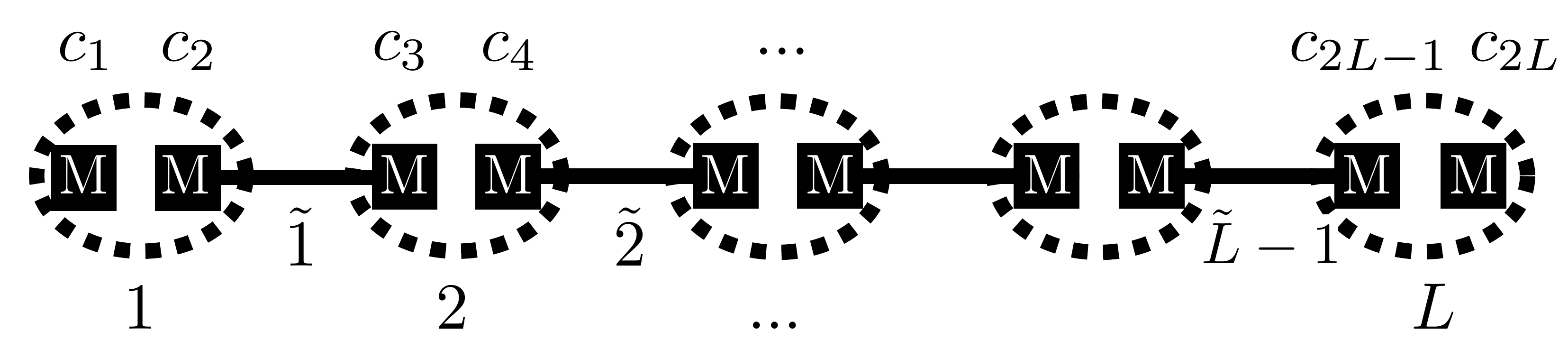}
  \caption{\label{fig:Kitaevchain} The Kitaev chain with $L$ sites and open boundary conditions. Each site $i$ (denoted by a dashed circle) can be occupied by one spinless fermion, and can be decomposed into two Majorana fermions (denoted by black dots) in $2i-1$ and $2i$. Associating the Majoranas $2i$ and $2i+1$ allows the construction of a new quasi local fermionic basis denoted with tildes. The ground states in the topological stereotypical regime will see its neighbouring Majorana fermions pairing up, so that, in the tilded basis, each site is unoccupied. Only the two Majorana on the edges do not need to pair up.}
\end{figure}

We give here a brief reminder of the Kitaev wire of Kitaev's seminal paper Ref.~\citen{Kitaev01} for the unfamiliar reader. The Kitaev wire is a chain of $L$ spinless fermions with open boundary conditions described by the Hamiltonian:
\begin{equation}
\begin{split}
H=&\sum_{j=1}^{L-1}\left(-t \left(a_j^\dagger a_{j+1}+a_{j+1}^\dagger a_{j}\right) \right. \\
&\left. -\mu \left(a_j^\dagger a_j-\frac{1}{2} \right)+ \Delta a_j a_{j+1}+ \Delta^* a_{j+1}^\dagger a_j^\dagger \right),
\end{split}
\label{eq:hamKitini}
\end{equation}
where $t$ is the hopping amplitude, $\mu$ is the chemical potential, and $\Delta= \lvert \Delta \rvert e^{i \theta}$ the induced superconducting gap. It is convenient to absorb the complex phase of the latter in a (completely local) redefinition of the local creation and annihilations operators $a_j^\dagger$ and $a_j$ such that:
\begin{equation}
(a_j^\dagger,a_j) \rightarrow (e^{-i \theta / 2 }a_j^\dagger,e^{i \theta / 2 }a_j), \label{eq:def_maj_ferm}
\end{equation}
and consider Eq.~\ref{eq:hamKitini} with $\Delta$ real only. It is then useful to introduce the Majorana fermions operators $c_j$ (for $j=1,...,L$):
\begin{equation}
c_{2j-1}= a_j +a_j^\dagger , \;\;\;\; c_{2j}=\frac{a_j - a_j^\dagger}{i},
\end{equation}
such that:
\begin{equation}
\lbrace c_m , c_l \rbrace = 2 \delta_{m,l}, \;\;\;\; c_m^\dagger = c_m,
\end{equation}
where $\delta_{m,l}$ is the Kronecker delta. The Hamiltonian Eq.~\ref{eq:hamKitini} then becomes:
\begin{equation}
\begin{split}
H=&\frac{i}{2}\sum_{j=1}^{L-1}\left( -\mu c_{2j-1}c_{2j}+ \left(t+\lvert \Delta \rvert \right)c_{2j}c_{2j+1} \right. \\
&\left. +\left(-t+\lvert \Delta \rvert \right) c_{2j-1}c_{2j+2}\right).
\end{split}
\label{eq:hamKitMaj}
\end{equation}
In the special regime of parameters when $\lvert \Delta \rvert = t  > 0 $ and $\mu=0$ (which we call the stereotypical regime), the Hamiltonian Eq.~\ref{eq:hamKitMaj} becomes: 
\begin{equation}
H=it\sum_{j=1}^{L-1}c_{2j}c_{2j+1},
\label{eq:hamKitMajspec}
\end{equation}
where it is important to note that the Majorana operators appearing in each term of the sum are not from the same sites. One can define new local fermionic creation and annihilation operators on the link such that (for $j=1,...,L-1$ only):
\begin{equation}
\tilde{a}_j=\frac{c_{2j}+ic_{2j+1}}{2}, \;\;\;\; \tilde{a}_j^\dagger =\frac{c_{2j}-ic_{2j+1}}{2},
\end{equation}
that only mixes two neighbouring sites. The Hamiltonian Eq.~\ref{eq:hamKitMajspec} becomes diagonal:
\begin{equation}
H=2t\sum_{j=1}^{L-1}\left(\tilde{a}_j^\dagger\tilde{a}_j-\frac{1}{2}\right),
\label{eq:hamKitMajspec2}
\end{equation}
and has two degenerate ground states, each pairing a Majorana fermion of one edge with a Majorana fermion of the other (cf Fig.~\ref{fig:Kitaevchain}). Defining the non-local operators:
\begin{equation}
b=\frac{c_{2L}+ic_1}{2}, \;\;\;\; b^\dagger=\frac{c_{2L}-ic_1}{2},
\end{equation}
the two ground states $\lvert 0 \rangle$ and $\lvert 1 \rangle$ satisfy:
\begin{subequations}
\begin{align}
\forall j \in \llbracket 1, L-1 \rrbracket, \;\; \tilde{a}_j \lvert 0 \rangle &= 0, \\
\forall j \in \llbracket 1, L-1 \rrbracket, \;\; \tilde{a}_j \lvert 1 \rangle &= 0,  \\
  b\lvert 0 \rangle &= 0, \\
 b^\dagger \lvert 0 \rangle &= \lvert 1 \rangle.
\end{align}
\label{eq:defannih}
\end{subequations}
In the case of periodic boundary conditions, $\vert 0 \rangle$ becomes the only ground state.
%%%%%%%%%%%%%%%%%%%%%%%%%%%%%%%%%%%%%%%%%%%%%%%%%%%%%%%%
%%%%%%%%%%%%%%%%%%%%%%%%%%%%%%%%%%%%%%%%%%%%%%%%%%%%%%%%

\subsection{The entanglement properties of the topological phase in the stereotypical regime}\label{sec:entprop}

\begin{figure}[htb]
  \centering
   \begin{overpic}[width=1.0\linewidth]{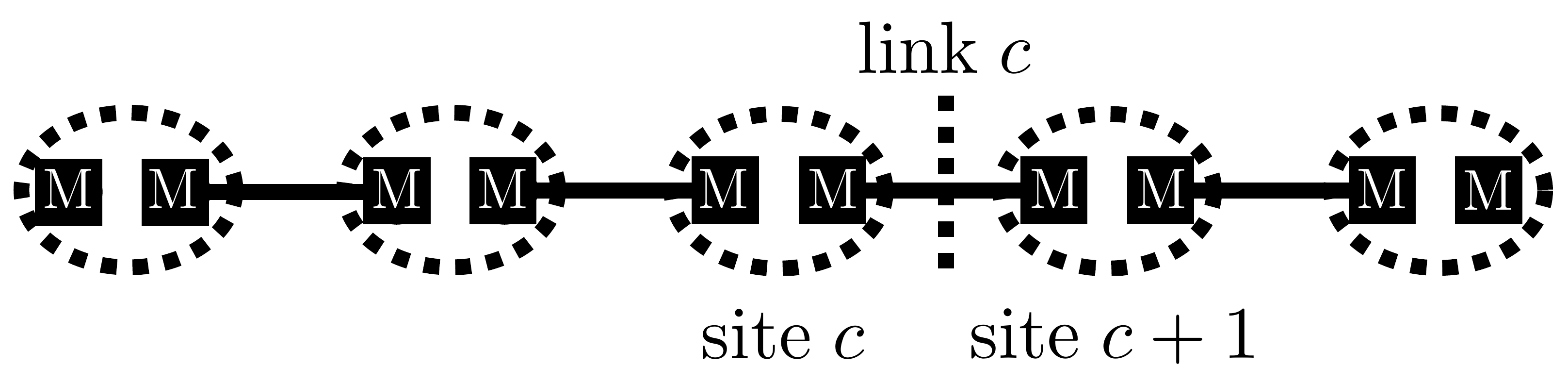}
  \put (28,19.5) {\Huge A} \put (78,19.5) {\Huge B}
\end{overpic}
  \caption{\label{fig:Kitaevchaincut} A physical cut can only be done between sites, here, on the link $c$, partitioning the chain into two subsets: $A$ and $B$.}
\end{figure}

To understand the entanglement properties of this topological phase and analytically compute the disconnected entanglement entropy $S^D$, it is useful to compute any reduced density matrices for the ground states $\lvert 0 \rangle $ and $\lvert 1 \rangle$ obtained in the stereotypical regime when $ \lvert \Delta \rvert =t > 0$ and $\mu=0$. To do so, it is useful to rewrite these states in the ``second quantization" formalism, but in the new tilted basis where:
\begin{subequations}
\begin{align}
\forall j \in \llbracket 1, L-1 \rrbracket, \;\;\tilde{n}_j &= \tilde{a}_j^\dagger \tilde{a}_j, \\
n_b &= b^\dagger b,
\end{align}
\end{subequations}
where the index $b$ stands for boundary. In that case, the two ground states can be rewritten as:
\begin{subequations}
\begin{align}
\lvert 0 \rangle &= \lvert \tilde{n}_1 =0, \tilde{n}_2 =0, \dots , \tilde{n}_{j-1}=0, n_b =0 \rangle, \\
\lvert 1 \rangle &= \lvert \tilde{n}_1 =0, \tilde{n}_2 =0, \dots , \tilde{n}_{j-1}=0, n_b =1 \rangle,
\end{align}
\label{eq:gsnewbasis}
\end{subequations}
which is a quasi local basis in the sense that each $\tilde{n}_j$ can be expressed in terms of operators acting only on sites $j$ and $j+1$. 
For a connected bipartition of the system, as illustrated in Fig.~\ref{fig:Kitaevchaincut}, the ground states [Eqs.~\ref{eq:gsnewbasis}]
cannot be written as a product state, due to both the presence of edge states and the way  Majorana fermions are linked in the bulk of the system (for instance, see link $c \in \llbracket 1, L-1 \rrbracket $ 
of Fig.~\ref{fig:Kitaevchaincut}). 

%%%%%%%%%%%%%%%%%%%%%%%%%%%%%%%%%%%%%%%%%%%%%%%%%%%%%%%%
%%%%%%%%%%%%%%%%%%%%%%%%%%%%%%%%%%%%%%%%%%%%%%%%%%%%%%%%

To properly do the partial trace and obtain the reduced density matrix $\rho_A$, it is  better to express the ground states in terms of a local basis for both $A$ and $B$. This becomes possible when rewriting the two parts of the open Kitaev wire as two open Kitaev wires connected into a singlet on the link $c$. Calling $L_A$ the size of $A$, and $L_B$ the size of $B$, such that $L_A + L_B = L$, we define a new fermionic basis, local in $A$ and $B$:
\begin{subequations}
\begin{align}
a_A &=\frac{1}{2}\left( c_{2L_A}+ic_1 \right), \\ a_A^\dagger &=\frac{1}{2}\left( c_{2L_A}-ic_1 \right), \\
a_B &=\frac{1}{2}\left( c_{2L_A+2L_B}+ic_{2L_A+1} \right), \\ a_B^\dagger &=\frac{1}{2}\left( c_{2L_A+2L_B}-ic_{2L_A+1} \right), \\
\tilde{a}_c &= \frac{1}{2}\left( c_{2L_A}+ic_{2L_A+1} \right), \\ \tilde{a}_c^\dagger &=\frac{1}{2}\left( c_{2L_A}-ic_{2L_A+1} \right), \\
b&=\frac{1}{2}\left(c_{2L}+ic_1\right), \\ b^\dagger &=\frac{1}{2}\left(c_{2L}-ic_1\right),
\end{align}
\label{eq:opnewbasis}
\end{subequations}
so that $a_A$ and $a_B$ (and hermitian conjugate) act as boundary operators for the subchains $A$ and $B$ respectively. Hence, in second quantization, and after dropping the redundant mentions of the $\tilde{n}_j$, $j\in \llbracket 1, L-1 \rrbracket \setminus \lbrace c=L_A \rbrace $, the two ground states of the full chain are $\lvert n_b=0, \tilde{n}_c=0 \rangle$ and $\lvert n_b=1, \tilde{n}_c=0 \rangle$. The local basis of $A$ and $B$ is $\lbrace \lvert n_A, n_B \rangle \rbrace$ where $n_A= a_A^\dagger a_A$ and $n_B= a_B^\dagger a_B$ take the values $0$ or $1$. Using Eqs.~\ref{eq:defannih} and Eqs.~\ref{eq:opnewbasis}, we find:
\begin{subequations}
\begin{align}
\lvert n_b=0, \tilde{n}_c=0 \rangle  &= -\frac{1}{\sqrt{2}}\left(\lvert n_A=1, n_B=0 \rangle \right. \nonumber\\
& \;\;\;\; \left.- \lvert n_A=0, n_B=1 \rangle \right), \\
&(= \lvert 0\rangle) \nonumber\\
\lvert n_b=1, \tilde{n}_c=1 \rangle &= \frac{1}{\sqrt{2}}\left(\lvert n_A=1, n_B=0 \rangle \right. \nonumber\\
& \;\;\;\; \left.+ \lvert n_A=0, n_B=1 \rangle \right), \\
\lvert n_b=1, \tilde{n}_c=0 \rangle &= \frac{1}{\sqrt{2}}\left(\lvert n_A=0, n_B=0 \rangle \right. \nonumber\\
& \;\;\;\; \left.+ \lvert n_A=1, n_B=1 \rangle \right), \\
&(= \lvert 1 \rangle\nonumber ) \\
\lvert n_b=0, \tilde{n}_c=1 \rangle &= \frac{1}{\sqrt{2}}\left(-\lvert n_A=0, n_B=0 \rangle \right. \nonumber\\
& \;\;\;\; \left.+ \lvert n_A=1, n_B=1 \rangle \right).
\end{align}
\label{eq:localbasis}
\end{subequations}
Tracing over $B$ is immediate, as the only vectors of the basis of $B$ with possible non zero contributions are $\lvert n_B =0 \rangle$ and $\lvert n_B =1 \rangle$. In particular:
\begin{equation}
\begin{split}
\rho_A \left( \lvert n_b=0, \tilde{n}_c=0 \rangle \langle n_b=0, \tilde{n}_c=0 \rvert \right)\;\;\;\;\;\; \;\;\;\;\;\;\;\;\;\;\\
= \frac{1}{2}\left( \lvert n_A=1 \rangle \langle n_A = 1 \rvert + \lvert n_A=0 \rangle \langle n_A = 0 \rvert \right),
\end{split}
\end{equation}
of entanglement entropy $S_A = \log 2$. The same happens for the other ground state.

Notice that the partial trace for fermions can induce a change of sign compared to the bosonic case. For example:
\begin{equation}
\begin{split}
\Tr_B \left( \lvert n_A=1, n_B= 1 \rangle \langle n_A=0 , n_B=1 \rvert \right) \;\;\;\;\;\; \\
 = - \lvert n_A = 1 \rangle \langle n_A=0 \rvert .
\end{split}
\end{equation}
Using Eqs.~\ref{eq:localbasis}, it is possible to get the expressions of the ground states in the local basis of an arbitrary partition. Additionally, taking a partition where all sites are their own individual subsets leads to the expression of the ground states in the original basis, up to a phase.
%%%%%%%%%%%%%%%%%%%%%%%%%%%%%%%%%%%%%%%%%%%%%%%%%%%%%%%%
\begin{figure}[htb]
   \includegraphics[width=1.0\linewidth]{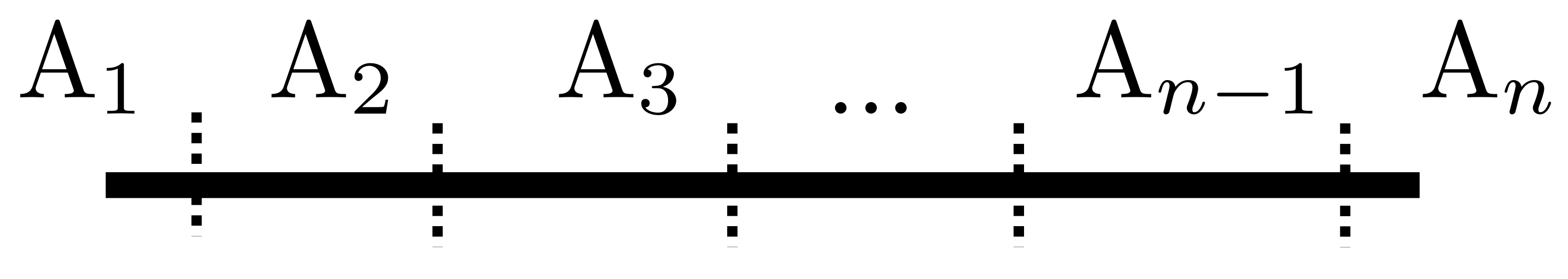}
  \caption{\label{fig:Kitaevchainmulticut} A partition of the chain into $n$ consecutive connected subsets $A_1$, $A_2$, ..., $A_n$.}
\end{figure} 
%%%%%%%%%%%%%%%%%%%%%%%%%%%%%%%%%%%%%%%%%%%%%%%%%%%%%%%%
The general expression of the reduced density matrix for an arbitrary partition of the system is obtained recursively, by considering the partition $A_1, A_2, \dots , A_n$ of connected subsets $A_i$ that are next to each others like in Fig.~\ref{fig:Kitaevchainmulticut}. 
Calling $A=A_1 $ and $B=B_1= \cup_{i=2}^{n}A_i$ allow use of Eqs,~\ref{eq:localbasis} to express the two ground states in the local basis of $A$ and $B$ instead of $A \cup B$.
The recurrence follows. Naming $c_i$ the link between subsets $A_i$ and $A_{i+1}$ and constructing the ``local boundary operators" $a_{A_i}$ and $a_{A_i}^\dagger$ for the subset $A_i$ and $a_{B_j}$ and $a_{B_j}^\dagger$ for $B_j=\cup_{i=j+1}^{n}A_i$ similarly to Eqs.~\ref{eq:localbasis}, the recurrence can be written as ($\forall n \geqslant2$):
\begin{subequations}
\begin{align}
&u_n\left(A_1,\dots,A_n \right)\\&\doteq\lvert n_b=0, \tilde{n}_{c_1}=0,\dots, \tilde{n}_{c_{n-1}}=0 \rangle (=\lvert 0 \rangle)\\
&=\frac{1}{\sqrt{2}}\left(\lvert n_{A_1}=0,n_{B_1}=1,\tilde{n}_{c_2}=0,\dots, \tilde{n}_{c_{n-1}}=0 \rangle \right. \nonumber \\
& \;\;\;\;\;\;\;\;\;\left. -\lvert n_{A_1}=1,n_{B_1}=0,\tilde{n}_{c_2}=0,\dots, \tilde{n}_{c_{n-1}}=0 \rangle \right)\\
&=\frac{1}{\sqrt{2}}\left(\lvert n_{A_1}=0\rangle \otimes v_{n-1}\left(A_2,\dots,A_n\right) \right. \nonumber \\
& \;\;\;\;\;\;\;\;\;\left. - \lvert n_{A_1}=1\rangle \otimes u_{n-1}\left(A_2,\dots,A_n\right)\right),\\
%%%%%%%%%%%%%%%%%%%
&v_n\left(A_1,\dots,A_n \right)\\&\doteq\lvert n_b=1, \tilde{n}_{c_1}=0,\dots, \tilde{n}_{c_{n-1}}=0 (=\lvert 1 \rangle)\rangle \\
&=\frac{1}{\sqrt{2}}\left(\lvert n_{A_1}=0,n_{B_1}=0,\tilde{n}_{c_2}=0,\dots, \tilde{n}_{c_{n-1}}=0 \rangle \right. \nonumber \\
& \;\;\;\;\;\;\;\;\; \left.+\lvert n_{A_1}=1,n_{B_1}=1,\tilde{n}_{c_2}=0,\dots, \tilde{n}_{c_{n-1}}=0 \rangle \right)\\
&=\frac{1}{\sqrt{2}}\left(\lvert n_{A_1}=0\rangle \otimes u_{n-1}\left(A_2,\dots,A_n\right) \right. \nonumber \\
& \;\;\;\;\;\;\;\;\; \left. + \lvert n_{A_1}=1\rangle \otimes v_{n-1}\left(A_2,\dots,A_n\right)\right).
\end{align}
\label{eq:rec1}
\end{subequations}
Calling:
\begin{subequations}
\begin{align}
U_n&=u_n+iv_n, \\
V_n&=u_n-iv_n, \\
\lvert +_j \rangle &=1/\sqrt{2}\left(i\lvert n_{A_j}=0\rangle -\lvert n_{A_j}=1\rangle\right),\\
\lvert -_j \rangle &=1/\sqrt{2}\left(-i\lvert n_{A_j}=0\rangle -\lvert n_{A_j}=1\rangle\right),
\end{align}
\label{eq:redef}
\end{subequations}
these relations become:
\begin{subequations}
\begin{align}
&U_n\left(A_1,\dots,A_n \right)=\lvert +_1 \rangle \otimes V_{n-1}\left(A_2,\dots,A_n \right) \\
&=\left\lbrace \begin{array}{cc}
\lvert +_1 -_2 \dots -_{n-2} \rangle \otimes U_2\left(A_{n-1},A_n \right) & \textrm{if n is even} \\
\lvert +_1 -_2 \dots +_{n-2} \rangle \otimes V_2\left(A_{n-1},A_n \right) & \textrm{if n is odd}
\end{array} \right.\\
%%%%%%%%%%%%%%%%%%%
&V_n\left(A_1,\dots,A_n \right)=\lvert -_1 \rangle \otimes U_{n-1}\left(A_2,\dots,A_n \right) \\
&=\left\lbrace \begin{array}{cc}
\lvert -_1 +_2 \dots +_{n-2} \rangle \otimes V_2\left(A_{n-1},A_n \right) & \textrm{if n is even} \\
\lvert -_1 +_2 \dots -_{n-2} \rangle \otimes U_2\left(A_{n-1},A_n \right) & \textrm{if n is odd}
\end{array} \right.
\end{align}
\label{eq:rec2}
\end{subequations}
where:
\begin{subequations}
\begin{align}
u_2\left(A_{n-1},A_{n}\right)&=\frac{1}{\sqrt{2}}\left(\lvert n_{A_{n-1}}=0,n_{A_{n}}=1 \rangle \right. \\
&\left. -\lvert n_{A_{n-1}}=1,n_{A_{n}}=0 \rangle\right), \nonumber\\
v_2\left(A_{n-1},A_{n}\right)&=\frac{1}{\sqrt{2}}\left(\lvert n_{A_{n-1}}=0,n_{A_{n}}=0 \rangle \right. \\
&\left. +\lvert n_{A_{n-1}}=1,n_{A_{n}}=1 \rangle\right), \nonumber
\end{align}
\label{eq:rec3}
\end{subequations}
so that:
\begin{subequations}
\begin{align}
U_2\left(A_{n-1},A_{n}\right)&=\sqrt{2}i\lvert +_{n-1}-{n} \rangle,\\
V_2\left(A_{n-1},A_{n}\right)&=\sqrt{2}i\lvert -_{n-1}+_{n} \rangle .
\end{align}
\label{eq:rec4}
\end{subequations}
Therefore, in a local basis of $A_1, \dots, A_n$, the ground states become:
\begin{subequations}
\begin{align}
\lvert n_b=0 \rangle &= u_n\left( A_1, \dots, A_n \right) \\
&=\frac{i}{\sqrt{2}}\left(\lvert +_1 -_2 \dots \rangle  + \lvert -_1 +_2 \dots \rangle\right), \\
\lvert n_b=1 \rangle &= v_n\left( A_1, \dots, A_n \right) \\
&=\frac{i}{\sqrt{2}}\left(\lvert +_1 -_2 \dots \rangle  - \lvert -_1 +_2 \dots \rangle\right).
 \end{align}
\label{eq:rec5}
\end{subequations}
These states are not N\'eel states because they are made out of fermions. It becomes clear in the basis of the subsets $\lbrace \otimes \lvert n_{A_j} \rangle \rbrace_j$ up to the global phase change:
\begin{equation}
\lvert \tilde{0}_j \rangle \doteq (-1)^j i \lvert 0_{A_j} \rangle, \;\;\;\; \mathrm{and} \;\;\;\; \lvert \tilde{1}_j \rangle \doteq \lvert 1_{A_j} \rangle.
\end{equation}
In that case:
\begin{subequations}
\begin{align}
\lvert +_1 -_2 \dots \rangle  &= \left(-\frac{1}{\sqrt{2}}\right)^n \otimes_{i=1}^n \left[\lvert \tilde{0}_i \rangle + \lvert \tilde{1}_i \rangle \right] \\
&= \left(-\frac{1}{\sqrt{2}}\right)^n \;\;\;\;\;\; \sum_{\mathclap{\lbrace n_{A_i} \rbrace_{i \in \llbracket 1,n \rrbracket} =0,1}} \lvert \lbrace n_{A_i} \rbrace_{i \in \llbracket 1,n \rrbracket} \rangle \\
\lvert -_1 +_2 \dots \rangle &= \left(-\frac{1}{\sqrt{2}} \right)^n \nonumber\\
&\;\;\;\;\times \sum_{\mathclap{\lbrace n_{A_i} \rbrace_{i \in \llbracket 1,n \rrbracket} =0,1}} (-1)^{n-\sum n_{A_i}}\lvert \lbrace n_{A_i} \rbrace_{i \in \llbracket 1,n \rrbracket} \rangle
\end{align}
\label{eq:rec6}
\end{subequations}
so that:
\begin{equation}
S_{A_1 \cup A_3 \cup A_5 \cup \dots}= \lfloor \frac{n+1}{2} \rfloor \log 2,
\end{equation}
where $\lfloor \dots \rfloor$ is the floor function, and $n$ the number of partition. This last equation proves the exact additivity of the entropy in this case independently of the position of the cuts, which, in addition to the non-nullity of the contribution of the individual subsets, ensure the non-nullity of $S^D$ for any superposition of the ground states. Indeed, for $n=4$:
\begin{equation}
S^D = S_{A_1 \cup A_2} + S_{A_2 \cup A_4}- S_{A_2}- S_{A_1 \cup A_2 \cup A_4},
\label{eq:sqtopo2}
\end{equation}
becomes the net contribution of one connected subset only: $\log 2$. 
Alternatively, it is the contribution of two cut Bell pairs of Majorana fermions. This result is valid for both the von Neumann and the R\'enyi entropies. 

\subsection{Equivalence of the $S^D_n$} \label{sec:equiv}

For 1D gapped systems, all $S^D_n$ can be used interchangeably, with $n\geq1$ and finite the index of the R\'enyi entropy: 
\begin{itemize}
\item The property of minimum value shared by all $S_n$ imply that if $S_{n_1}\neq 0$, then $S_{n_2}\neq 0$ and vice-versa, so both are simultaneously non-zero.
\item The property of monotonicity and the fact that all gapped 1D phases have finite von Neumann entanglement entropy imposes all all $S_n$ to also be finite for 1D gapped systems, so no $S^D_n$ can diverge.
\item For a given $n$-entropy Translation invariance further imposes that all connected entropy will have the same value for big subset size, \textit{i.e.} if $A_{i+t}$ is the subset translated from $A_i$, then $S_n(A_{i+t})=S_n(A_{i})$ up to finite size effects.
\item The property of additivity is shared by all $S_n$.
\end{itemize}

\subsection{Comparison between the topological and non-topological phases away from the phase transition}

In the topological superconductor phase, away from any phase transition, both the von Neuman and R\'enyi entanglement entropies for large enough partitions (i.e., $L_A, L_B, L_D \gg \xi$ for the definitions of Fig.~\ref{fig:cuts} (a), or $L_{A_i} \gg \xi$ for all $i$ for Eq.~\ref{eq:sqtopo2}) are non zero and additive as 
demonstrated above in the stereotypical regime (in addition, see next section).
If A is a simply connected subset of a partition of the chain (independent of its position), then, $S_A=2 \Gamma$, where $\Gamma= (\log 2)/2$ is the contribution of one (Majorana) Bell pair. Eq.~\ref{eq:sqtopo2} then gives:
\begin{equation}
S^D=2 \Gamma + 2 \times 2 \Gamma -2 \Gamma - 2 \Gamma = 2 \Gamma
\end{equation}
In that regard, $S^D$ is not unique: combination such as $S_{A_1 \cup A_3}+S_{A_2 \cup A_4}-S_{A_1 \cup A_4}-S_{A_2 \cup A_3}$ would have also work as detectors, 
but with less experimental relevance, and a more complicated interpretation in terms of mutual information. 
The quantitative equality $S^D=S^D_2$ is here coincidental and is not a generic feature for other systems.

For the band insulator phase and for large enough partitions, the contribution of each term of Eq.~\ref{eq:sqtopo2}
is proportional to the number of boundaries of the partition, as we expect from the so called \textit{area law}.
For the 1D OBC systems, one can argue that $S^D = 0$
by simply counting the number of boundaries of the partitions shown in Fig.~\ref{fig:cuts} a), which gives:
\begin{equation}
S^D= \Theta + 3 \Theta -2 \Theta - 2 \Theta = 0,
\end{equation}
where
$\Theta$ is the contribution for the entanglement of a single boundary.
While $\Theta$ is model-parameter depending, 
the ratios between each terms is constant in the limit $L_A, L_B, L_D \gg \xi$, 
ensuring $S^D = 0$. In the limit $\mu \to \infty$, each term is identically zero.
The result $S^D = 0$ in the non-topological phase is backed up by numerical simulations.

We also discuss the behavior of $S^{D}$ for other non-toplogical phases.
For the case of a gapped phase displaying ground state equivalent to \textit{e.g.} a maximally entangled N\'eel state, the entanglement entropy becomes non zero for each term, but is not additive, 
such that (calling $\Theta$ the contribution of the only Bell pair of spin 1/2)~:
\begin{equation}
S^D=\Theta +\Theta -\Theta -\Theta = 0.
\end{equation}

For a critical phase, conformal field theory predicts a vanishing contribution. This is in agreement with all of our microscopic simulations. For critical, non-conformal points, we are not aware of any field theoretical prediction.

Finally, for the case of a rigorously dimerized phase, $S^D$ is well-defined and can be considered additive, but is not translation invariant. More precisely, let us define $\epsilon_{ij}$, where $i,j=1,2,3,4$, such that $\epsilon_{ij}=1$ if the cut between $A_i$ and $A_j$ is on a dimer, and $\epsilon_{ij}=0$ otherwise. Then, if $\Theta$ is the contribution of one dimer:
\begin{equation}
\begin{split}
S^D &=\Theta\left(\epsilon_{41}+\epsilon_{23}+\epsilon_{12}+\epsilon_{23}+\epsilon_{34}+\epsilon_{41} \right.
\\ & \;\;\;\;\;\;\;\;\;\left.-\epsilon_{12}-\epsilon_{23}-\epsilon_{23}-\epsilon_{34} \right) \\
&=2\Theta\epsilon_{41},
\end{split}
\end{equation}
which is always zero if $A_1$ and $A_4$ are subsets at both ends of the open chain. $S^D$ is not translation invariant in the case of periodic boundary conditions.

%%%%%%%%%%%%%%%%%%%%%%%%%%%%%%%%%%%%%%%%%%%%%%%%%%%%%%%%
\section{Additional information on the numerical methods}
\subsection{Computation using free-fermion correlation functions at equilibrium} \label{sec:ff}
The main challenge of the numerical analysis consists in calculating the four von Neumann entanglement entropies that are included in the definition of the disconnected topological entanglement entropy $S^{D}$ of Eq.~\ref{eq:sqtopo}. We recall that each of these four quantities is associated with a specific possibly disconnected bipartition of the chain, following the scheme shown in Fig.~\ref{fig:cuts} a).

The starting point for evaluating von Neumann or R\'enyi entanglement entropy $S_{A}$ of a generic bipartition $A$ and $B$ (not necessarily simply connected) is the computation of the reduced density matrix $\rho_{A}=\mathrm{T\mathrm{r}_{B}\left|\psi\right\rangle \left\langle \psi\right|}$ where $\left|\psi\right\rangle $ is a ground state of the whole system. For the Kitaev wire without interactions, \textit{i.e.} when $U=0$, the Hamiltonian Eq.~\ref{eq:hamKitini} is a free-fermion Hamiltonian that only conserves fermion number parity. It can be diagonalized by a Bogoliubov transformation. The reduced density matrix $\rho_{A}$ of a generic partition can be interpreted as the thermal density matrix at temperature $T=1$ for an entanglement Hamiltonian $\mathcal{H}_{A}$. Then $\rho_{A}=Z_{A}^{-1}e^{-\mathcal{H_{A}}}$ with $Z_{A}=\mathrm{Tr}\left[e^{-\mathcal{H_{A}}}\right]$ \cite{Peschel01, Peschel02, Peschel03}. It is then possible to compute $\rho_{A}$ of a generic partition by following the well-established approach of Ref.~\citen{Peschel04}. 
The first step of this very general procedure is the computation of the correlation matrices in the original ground-state: $C_{nm}=\left\langle \psi\left|a_{n}^{\dagger}a_{m}\right|\psi\right\rangle$ and $F_{nm}=\left\langle \psi\left|a_{n}^{\dagger}a_{m}^{\dagger}\right|\psi\right\rangle$ where $n$ and $m$ run over the sites of the subset $A$. At equilibrium, $F_{nm}$ is real, so that the spectrum of the entanglement Hamiltonian can be determined by numerically solving the eigenvalue problem~\cite{Peschel04}:
\begin{equation}
\left(2\hat{C}-2\hat{F}-1\right)\left(2\hat{C}+2\hat{F}-1\right)\phi_{l}=\tanh^{2}\left(\frac{\epsilon_{l}}{2}\right)\phi_{l}.
\label{eq:free_fermion_eigenvalue_problem}
\end{equation}
$\epsilon_{l}$ are the eigenvalues of the entanglement Hamiltonian with eigenvectors $\phi_{l}$. Once the spectrum $\epsilon_{l}$ is obtained, we can easily compute the reduced density matrix $\rho_{A}$ and the entanglement entropy of any bipartition, connected or not.

We carried out these numerical calculations with arbitrary precision by using \emph{mpmath} Python library~\cite{mpmath_library}. Two requirements arise: avoiding numerical precision problems when the eigenvalues of the left side of Eq.~\ref{eq:free_fermion_eigenvalue_problem} are approximately $0$ and avoiding divergence problems when they are very close to $1$. These issues are fixed by taking a number of digits proportional to the total size of the system $L$: we set mp.dps = $20\times L$ (number of digits in the Python library). While the approach allows generation of a lot of data with relatively little cost, the interaction case, as well as the bilinear biquadratic model remain inaccessible with this algorithm. Instead, we have to switch to the Density Matrix Product State (DMRG) technique presented Sec.~\ref{sec:DMRG}.

\subsection{Sudden quenches using free-fermion correlation functions}
We check that $S^D$ is a topological invariant by looking at its time evolution after sudden quenches. The system starts in its ground state before the quench, for a given set of parameters of the Hamiltonian. At $t=0$ we change the value of one of these parameters. We then let the system evolve in time.

The generic interacting case is very challenging to follow because the interacting term induces a time evolution of an extensive number of eigenstates of the spectrum. For the free-fermion case, the same previous numerical approach of Sec.~\ref{sec:ff} still applies with the same efficiency. The matrix $F_{nm}$ is however complex during the time evolution of the system so that Eq.~\ref{eq:free_fermion_eigenvalue_problem} is not valid anymore. We follow instead Ref.~\citen{Peschel09} and use the definition of the Majorana fermions of Eq.~\ref{eq:def_maj_ferm}. The relevant $2L \times 2L$ correlation matrix is this time $M_{nm}=\langle c_nc_m \rangle$, rewritten as $M_{nm}=\delta_{nm}+\Gamma_{nm}$ with $n,m$ the (half-)site indices spanning only the relevant subset studied. The eigenvalues of $\Gamma_{nm}$ are $\pm \tanh{\varepsilon_l/2}$, with $\varepsilon_l$ the entanglement energies of the reduced density matrix associated with the subset.
The time dependent matrix $\Gamma_{nm}$ is directly linked to the time dependent Hamiltonian analytically, and then diagonalized numerically, similarly to Ref.~\citen{Fagotti08}.

We performed several quenches of different amplitudes inside both the TSC and the band insulator, as well as across the phase transition, as seen in Fig.~\ref{fig:quenches}. $S^D$ is plotted as a function of time for different chain lengths $L$ with $L/L_A=4$, $L_A=L_B$, using the same definition of Fig.~\ref{fig:cuts} a) for the partitions. \textit{a), b)} and \textit{c)} exhibits a quantized finite plateau depending on $L_A$. We define the time-size of the plateau with $t_c$: the time at which $S^D$ deviates from $\log 2$ to $0.95\log 2$. $t_c^\prime$ is time of revival, when $S^D$ increases from 0 to $0.05\log 2$. We plot $1/t_c$ or $4/t_c^\prime$ in \textit{d), e)} and \textit{f)} as a function of $1/L$.  The linear extrapolations of the length of the plateau show divergences of both time-size when $L \rightarrow \infty$ for each kind of quench: $S^D$ stays quantized in the thermodynamical limit because the drops/revivals are only finite size effects. Thus, $S^D$ does not change under coherent time evolution and behaves as a topological invariant.\\
We used the same implementation as in Sec.~\ref{sec:ff} with mp.dps$=3\times L$, enough to get reliable results.

\subsection{Computation using DMRG}\label{sec:DMRG}

For the case of the interacting Kitaev wire with $U\neq0$ (Eq.~\ref{eq:hamKitint}), the numerical study is now carried out with DMRG algorithm formulated in the matrix product state (MPS) language \cite{DMRG_Schollwoeck}.

\begin{figure}
\hfill \begin{overpic}[width=0.95\linewidth]{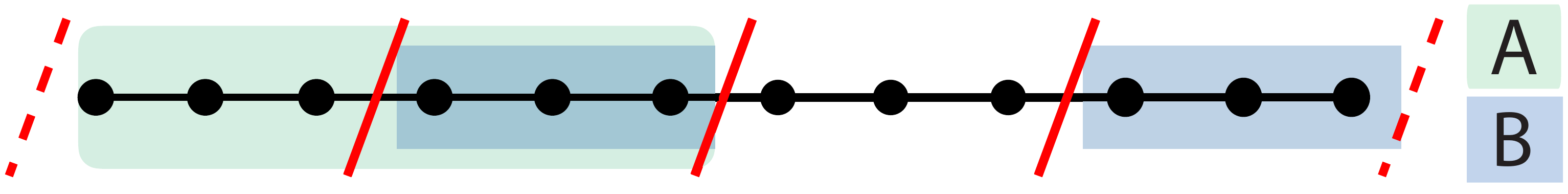}
\end{overpic}
\begin{overpic}[width=1.\linewidth]{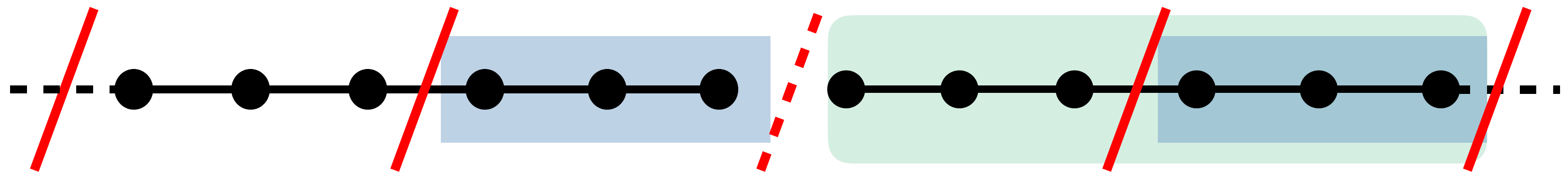}
\put (0,9) {(i)}
\end{overpic}
\begin{overpic}[width=1.\linewidth]{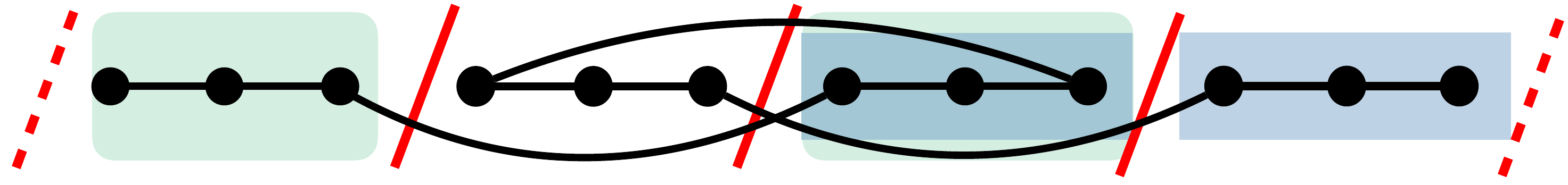}
\put (0,9) {(ii)}
\end{overpic}
\caption{\label{fig:Scheme_of_Partitions}Scheme of the partitions. Two ``tricks" are used to compute only standard bipartition of the chain: (i) a circular permutation of the partition used for $S(\rho_{A}$), $S(\rho_{A\cup B})$ and $S(\rho_{A\cap B})$ and (ii) re-indexing the sites, thus making the Hamiltonian long-range, but making the subset $B$ connected and used for $S(\rho_{B})$.}
\end{figure}

The main challenge is again the computation of von Neumann or R\'enyi entanglement entropies of disconnected bipartitions of the chain in order to determine the behavior of $S^{D}$ of Eq.~\ref{eq:sqtopo}. For DMRG however, only the standard connected bipartition of the MPS state into two halves is very efficient. This efficiency comes from the intimate structure of an MPS state and the use of left/right-orthogonality condition \cite{DMRG_Schollwoeck}. The MPS tensors themselves already give the eigenstates of the reduced density matrix of a block of sites starting from either the left or the right edge of the system. Instead, for generic bipartitions of the chain, such as partition $B$, $A\cup B$ or $A\cap B$ of Fig.~\ref{fig:cuts} a) the calculation of the reduced density matrix is highly non-trivial as it involves several multi-index tensor contractions. As a result, the computational cost
scales exponentially with the size of the subset.\\
We circumvent this numerical problem by reordering the MPS sites appropriately, thus considering a long-range Hamiltonian reproducing the original model. Two rearrangements are necessary, as shown in Fig. \ref{fig:Scheme_of_Partitions}. $(i)$ is a circular permutation of the original multi-partitions to efficiently calculate $S(\rho_{A}$), $S(\rho_{A\cup B})$ and $S(\rho_{A\cap B})$ because all these quantities are now related to right-most bipartitions of the chain. In the process, we only re-index the sites while their respective connectivity is left unchanged. $(ii)$ allows computation of $S(\rho_{B})$. In this way, we efficiently obtain the four entanglement entropies composing $S^{D}$.\\
In our numerical analysis with DMRG algorithm, we used a bond dimension up to $1200$, a truncation error of $10^{-8}$ and at least $30$ sweeps to reach the convergence and to ensure stability of our findings. We noticed that along the Peschel-Emery line of Ref.~\citen{Katsura15} in the TSC phase, $S^{D}$ converge even faster to its quantized value of $\log 2$.

\phantomsection
\addcontentsline{toc}{chapter}{Bibliography}
\bibliography{biblioentang}

\begin{thebibliography}{63}%
\makeatletter
\providecommand \@ifxundefined [1]{%
 \@ifx{#1\undefined}
}%
\providecommand \@ifnum [1]{%
 \ifnum #1\expandafter \@firstoftwo
 \else \expandafter \@secondoftwo
 \fi
}%
\providecommand \@ifx [1]{%
 \ifx #1\expandafter \@firstoftwo
 \else \expandafter \@secondoftwo
 \fi
}%
\providecommand \natexlab [1]{#1}%
\providecommand \enquote  [1]{``#1''}%
\providecommand \bibnamefont  [1]{#1}%
\providecommand \bibfnamefont [1]{#1}%
\providecommand \citenamefont [1]{#1}%
\providecommand \href@noop [0]{\@secondoftwo}%
\providecommand \href [0]{\begingroup \@sanitize@url \@href}%
\providecommand \@href[1]{\@@startlink{#1}\@@href}%
\providecommand \@@href[1]{\endgroup#1\@@endlink}%
\providecommand \@sanitize@url [0]{\catcode `\\12\catcode `\$12\catcode
  `\&12\catcode `\#12\catcode `\^12\catcode `\_12\catcode `\%12\relax}%
\providecommand \@@startlink[1]{}%
\providecommand \@@endlink[0]{}%
\providecommand \url  [0]{\begingroup\@sanitize@url \@url }%
\providecommand \@url [1]{\endgroup\@href {#1}{\urlprefix }}%
\providecommand \urlprefix  [0]{URL }%
\providecommand \Eprint [0]{\href }%
\providecommand \doibase [0]{http://dx.doi.org/}%
\providecommand \selectlanguage [0]{\@gobble}%
\providecommand \bibinfo  [0]{\@secondoftwo}%
\providecommand \bibfield  [0]{\@secondoftwo}%
\providecommand \translation [1]{[#1]}%
\providecommand \BibitemOpen [0]{}%
\providecommand \bibitemStop [0]{}%
\providecommand \bibitemNoStop [0]{.\EOS\space}%
\providecommand \EOS [0]{\spacefactor3000\relax}%
\providecommand \BibitemShut  [1]{\csname bibitem#1\endcsname}%
\let\auto@bib@innerbib\@empty
%</preamble>
\bibitem [{\citenamefont {Amico}\ \emph {et~al.}(2008)\citenamefont {Amico},
  \citenamefont {Fazio}, \citenamefont {Osterloh},\ and\ \citenamefont
  {Vedral}}]{Amico2008}%
  \BibitemOpen
  \bibfield  {author} {\bibinfo {author} {\bibfnamefont {L.}~\bibnamefont
  {Amico}}, \bibinfo {author} {\bibfnamefont {R.}~\bibnamefont {Fazio}},
  \bibinfo {author} {\bibfnamefont {A.}~\bibnamefont {Osterloh}}, \ and\
  \bibinfo {author} {\bibfnamefont {V.}~\bibnamefont {Vedral}},\ }\href
  {\doibase 10.1103/RevModPhys.80.517} {\bibfield  {journal} {\bibinfo
  {journal} {Rev. Mod. Phys.}\ }\textbf {\bibinfo {volume} {80}},\ \bibinfo
  {pages} {517} (\bibinfo {year} {2008})}\BibitemShut {NoStop}%
\bibitem [{\citenamefont {Calabrese}\ and\ \citenamefont
  {Cardy}(2009)}]{Calabrese_2009}%
  \BibitemOpen
  \bibfield  {author} {\bibinfo {author} {\bibfnamefont {P.}~\bibnamefont
  {Calabrese}}\ and\ \bibinfo {author} {\bibfnamefont {J.}~\bibnamefont
  {Cardy}},\ }\href {\doibase 10.1088/1751-8113/42/50/504005} {\bibfield
  {journal} {\bibinfo  {journal} {J. Phys. A: Math. Theor.}\ }\textbf {\bibinfo
  {volume} {42}},\ \bibinfo {pages} {504005} (\bibinfo {year}
  {2009})}\BibitemShut {NoStop}%
\bibitem [{\citenamefont {Eisert}\ \emph {et~al.}(2010)\citenamefont {Eisert},
  \citenamefont {Cramer},\ and\ \citenamefont {Plenio}}]{Eisert2010}%
  \BibitemOpen
  \bibfield  {author} {\bibinfo {author} {\bibfnamefont {J.}~\bibnamefont
  {Eisert}}, \bibinfo {author} {\bibfnamefont {M.}~\bibnamefont {Cramer}}, \
  and\ \bibinfo {author} {\bibfnamefont {M.~B.}\ \bibnamefont {Plenio}},\
  }\href {\doibase 10.1103/RevModPhys.82.277} {\bibfield  {journal} {\bibinfo
  {journal} {Rev. Mod. Phys.}\ }\textbf {\bibinfo {volume} {82}},\ \bibinfo
  {pages} {277} (\bibinfo {year} {2010})}\BibitemShut {NoStop}%
\bibitem [{\citenamefont {Fradkin}(2013)}]{fradkinbook}%
  \BibitemOpen
  \bibfield  {author} {\bibinfo {author} {\bibfnamefont {E.}~\bibnamefont
  {Fradkin}},\ }\href@noop {} {\emph {\bibinfo {title} {{Field Theories of
  Condensed Matter Systems}}}}\ (\bibinfo  {publisher} {Cambridge University
  Press},\ \bibinfo {year} {2013})\BibitemShut {NoStop}%
\bibitem [{\citenamefont {Hamma}\ \emph {et~al.}(2005)\citenamefont {Hamma},
  \citenamefont {Ionicioiu},\ and\ \citenamefont {Zanardi}}]{Hamma_2005}%
  \BibitemOpen
  \bibfield  {author} {\bibinfo {author} {\bibfnamefont {A.}~\bibnamefont
  {Hamma}}, \bibinfo {author} {\bibfnamefont {R.}~\bibnamefont {Ionicioiu}}, \
  and\ \bibinfo {author} {\bibfnamefont {P.}~\bibnamefont {Zanardi}},\ }\href
  {\doibase 10.1016/j.physleta.2005.01.060} {\bibfield  {journal} {\bibinfo
  {journal} {Phys. Lett. A}\ }\textbf {\bibinfo {volume} {337}},\ \bibinfo
  {pages} {22} (\bibinfo {year} {2005})}\BibitemShut {NoStop}%
\bibitem [{\citenamefont {{Levin}}\ and\ \citenamefont
  {{Wen}}(2006)}]{2006PhRvL..96k0405L}%
  \BibitemOpen
  \bibfield  {author} {\bibinfo {author} {\bibfnamefont {M.}~\bibnamefont
  {{Levin}}}\ and\ \bibinfo {author} {\bibfnamefont {X.-G.}\ \bibnamefont
  {{Wen}}},\ }\href {\doibase 10.1103/PhysRevLett.96.110405} {\bibfield
  {journal} {\bibinfo  {journal} {\prl}\ }\textbf {\bibinfo {volume} {96}},\
  \bibinfo {eid} {110405} (\bibinfo {year} {2006})}\BibitemShut {NoStop}%
\bibitem [{\citenamefont {Kitaev}\ and\ \citenamefont
  {Preskill}(2006)}]{Kitaev_2006}%
  \BibitemOpen
  \bibfield  {author} {\bibinfo {author} {\bibfnamefont {A.}~\bibnamefont
  {Kitaev}}\ and\ \bibinfo {author} {\bibfnamefont {J.}~\bibnamefont
  {Preskill}},\ }\href {\doibase 10.1103/PhysRevLett.96.110404} {\bibfield
  {journal} {\bibinfo  {journal} {Phys. Rev. Lett.}\ }\textbf {\bibinfo
  {volume} {96}},\ \bibinfo {pages} {110404} (\bibinfo {year}
  {2006})}\BibitemShut {NoStop}%
\bibitem [{\citenamefont {Wen}(2019)}]{Wen19}%
  \BibitemOpen
  \bibfield  {author} {\bibinfo {author} {\bibfnamefont {X.-G.}\ \bibnamefont
  {Wen}},\ }\href {\doibase 10.1126/science.aal3099} {\bibfield  {journal}
  {\bibinfo  {journal} {Science}\ }\textbf {\bibinfo {volume} {363}},\ \bibinfo
  {pages} {3099} (\bibinfo {year} {2019})}\BibitemShut {NoStop}%
\bibitem [{\citenamefont {Depenbrock}\ \emph {et~al.}(2012)\citenamefont
  {Depenbrock}, \citenamefont {McCulloch},\ and\ \citenamefont
  {Schollw\"ock}}]{Depenbrock:2012aa}%
  \BibitemOpen
  \bibfield  {author} {\bibinfo {author} {\bibfnamefont {S.}~\bibnamefont
  {Depenbrock}}, \bibinfo {author} {\bibfnamefont {I.~P.}\ \bibnamefont
  {McCulloch}}, \ and\ \bibinfo {author} {\bibfnamefont {U.}~\bibnamefont
  {Schollw\"ock}},\ }\href {\doibase 10.1103/PhysRevLett.109.067201} {\bibfield
   {journal} {\bibinfo  {journal} {Phys. Rev. Lett.}\ }\textbf {\bibinfo
  {volume} {109}},\ \bibinfo {pages} {067201} (\bibinfo {year}
  {2012})}\BibitemShut {NoStop}%
\bibitem [{\citenamefont {Jiang}\ \emph {et~al.}(2012)\citenamefont {Jiang},
  \citenamefont {Wang},\ and\ \citenamefont {Balents}}]{Jiang:2012aa}%
  \BibitemOpen
  \bibfield  {author} {\bibinfo {author} {\bibfnamefont {H.-C.}\ \bibnamefont
  {Jiang}}, \bibinfo {author} {\bibfnamefont {Z.}~\bibnamefont {Wang}}, \ and\
  \bibinfo {author} {\bibfnamefont {L.}~\bibnamefont {Balents}},\ }\href
  {https://doi.org/10.1038/nphys2465} {\bibfield  {journal} {\bibinfo
  {journal} {Nat. Phys.}\ }\textbf {\bibinfo {volume} {8}},\ \bibinfo {pages}
  {902} (\bibinfo {year} {2012})}\BibitemShut {NoStop}%
\bibitem [{\citenamefont {Isakov}\ \emph {et~al.}(2011)\citenamefont {Isakov},
  \citenamefont {Hastings},\ and\ \citenamefont {Melko}}]{Isakov:2011aa}%
  \BibitemOpen
  \bibfield  {author} {\bibinfo {author} {\bibfnamefont {S.~V.}\ \bibnamefont
  {Isakov}}, \bibinfo {author} {\bibfnamefont {M.~B.}\ \bibnamefont
  {Hastings}}, \ and\ \bibinfo {author} {\bibfnamefont {R.~G.}\ \bibnamefont
  {Melko}},\ }\href@noop {} {\bibfield  {journal} {\bibinfo  {journal} {Nature
  Physics}\ }\textbf {\bibinfo {volume} {7}},\ \bibinfo {pages} {772} (\bibinfo
  {year} {2011})}\BibitemShut {NoStop}%
\bibitem [{\citenamefont {Abanin}\ and\ \citenamefont
  {Demler}(2012)}]{Abanin:2012aa}%
  \BibitemOpen
  \bibfield  {author} {\bibinfo {author} {\bibfnamefont {D.~A.}\ \bibnamefont
  {Abanin}}\ and\ \bibinfo {author} {\bibfnamefont {E.}~\bibnamefont
  {Demler}},\ }\href {https://doi.org/10.1103/physrevlett.109.020504}
  {\bibfield  {journal} {\bibinfo  {journal} {Phys. Rev. Lett.}\ }\textbf
  {\bibinfo {volume} {109}},\ \bibinfo {pages} {020504} (\bibinfo {year}
  {2012})}\BibitemShut {NoStop}%
\bibitem [{\citenamefont {Daley}\ \emph {et~al.}(2012)\citenamefont {Daley},
  \citenamefont {Pichler}, \citenamefont {Schachenmayer},\ and\ \citenamefont
  {Zoller}}]{Daley_2012}%
  \BibitemOpen
  \bibfield  {author} {\bibinfo {author} {\bibfnamefont {A.~J.}\ \bibnamefont
  {Daley}}, \bibinfo {author} {\bibfnamefont {H.}~\bibnamefont {Pichler}},
  \bibinfo {author} {\bibfnamefont {J.}~\bibnamefont {Schachenmayer}}, \ and\
  \bibinfo {author} {\bibfnamefont {P.}~\bibnamefont {Zoller}},\ }\href
  {\doibase 10.1103/PhysRevLett.109.020505} {\bibfield  {journal} {\bibinfo
  {journal} {Phys. Rev. Lett.}\ }\textbf {\bibinfo {volume} {109}},\ \bibinfo
  {pages} {020505} (\bibinfo {year} {2012})}\BibitemShut {NoStop}%
\bibitem [{\citenamefont {Elben}\ \emph {et~al.}(2018)\citenamefont {Elben},
  \citenamefont {Vermersch}, \citenamefont {Dalmonte}, \citenamefont {Cirac},\
  and\ \citenamefont {Zoller}}]{Elben_2018}%
  \BibitemOpen
  \bibfield  {author} {\bibinfo {author} {\bibfnamefont {A.}~\bibnamefont
  {Elben}}, \bibinfo {author} {\bibfnamefont {B.}~\bibnamefont {Vermersch}},
  \bibinfo {author} {\bibfnamefont {M.}~\bibnamefont {Dalmonte}}, \bibinfo
  {author} {\bibfnamefont {J.~I.}\ \bibnamefont {Cirac}}, \ and\ \bibinfo
  {author} {\bibfnamefont {P.}~\bibnamefont {Zoller}},\ }\href {\doibase
  10.1103/PhysRevLett.120.050406} {\bibfield  {journal} {\bibinfo  {journal}
  {Phys. Rev. Lett.}\ }\textbf {\bibinfo {volume} {120}},\ \bibinfo {pages}
  {050406} (\bibinfo {year} {2018})}\BibitemShut {NoStop}%
\bibitem [{\citenamefont {Vermersch}\ \emph {et~al.}(2018)\citenamefont
  {Vermersch}, \citenamefont {Elben}, \citenamefont {Dalmonte}, \citenamefont
  {Cirac},\ and\ \citenamefont {Zoller}}]{Vermersch_2018}%
  \BibitemOpen
  \bibfield  {author} {\bibinfo {author} {\bibfnamefont {B.}~\bibnamefont
  {Vermersch}}, \bibinfo {author} {\bibfnamefont {A.}~\bibnamefont {Elben}},
  \bibinfo {author} {\bibfnamefont {M.}~\bibnamefont {Dalmonte}}, \bibinfo
  {author} {\bibfnamefont {J.~I.}\ \bibnamefont {Cirac}}, \ and\ \bibinfo
  {author} {\bibfnamefont {P.}~\bibnamefont {Zoller}},\ }\href {\doibase
  10.1103/PhysRevA.97.023604} {\bibfield  {journal} {\bibinfo  {journal} {Phys.
  Rev. A}\ }\textbf {\bibinfo {volume} {97}},\ \bibinfo {pages} {023604}
  (\bibinfo {year} {2018})}\BibitemShut {NoStop}%
\bibitem [{\citenamefont {Pichler}\ \emph {et~al.}(2016)\citenamefont
  {Pichler}, \citenamefont {Zhu}, \citenamefont {Seif}, \citenamefont
  {Zoller},\ and\ \citenamefont {Hafezi}}]{Pichler:2016aa}%
  \BibitemOpen
  \bibfield  {author} {\bibinfo {author} {\bibfnamefont {H.}~\bibnamefont
  {Pichler}}, \bibinfo {author} {\bibfnamefont {G.}~\bibnamefont {Zhu}},
  \bibinfo {author} {\bibfnamefont {A.}~\bibnamefont {Seif}}, \bibinfo {author}
  {\bibfnamefont {P.}~\bibnamefont {Zoller}}, \ and\ \bibinfo {author}
  {\bibfnamefont {M.}~\bibnamefont {Hafezi}},\ }\href
  {https://doi.org/10.1103/PhysRevX.6.041033} {\bibfield  {journal} {\bibinfo
  {journal} {Phys. Rev. X}\ }\textbf {\bibinfo {volume} {6}},\ \bibinfo {pages}
  {041033} (\bibinfo {year} {2016})}\BibitemShut {NoStop}%
\bibitem [{\citenamefont {Dalmonte}\ \emph {et~al.}(2018)\citenamefont
  {Dalmonte}, \citenamefont {Vermersch},\ and\ \citenamefont
  {Zoller}}]{Dalmonte:2017aa}%
  \BibitemOpen
  \bibfield  {author} {\bibinfo {author} {\bibfnamefont {M.}~\bibnamefont
  {Dalmonte}}, \bibinfo {author} {\bibfnamefont {B.}~\bibnamefont {Vermersch}},
  \ and\ \bibinfo {author} {\bibfnamefont {P.}~\bibnamefont {Zoller}},\ }\href
  {https://arxiv.org/pdf/1707.04455} {\bibfield  {journal} {\bibinfo  {journal}
  {Nat. Phys.}\ }\textbf {\bibinfo {volume} {14}},\ \bibinfo {pages} {827}
  (\bibinfo {year} {2018})}\BibitemShut {NoStop}%
\bibitem [{\citenamefont {Cornfeld}\ \emph {et~al.}(2019)\citenamefont
  {Cornfeld}, \citenamefont {Sela},\ and\ \citenamefont
  {Goldstein}}]{Cornfeld_2019}%
  \BibitemOpen
  \bibfield  {author} {\bibinfo {author} {\bibfnamefont {E.}~\bibnamefont
  {Cornfeld}}, \bibinfo {author} {\bibfnamefont {E.}~\bibnamefont {Sela}}, \
  and\ \bibinfo {author} {\bibfnamefont {M.}~\bibnamefont {Goldstein}},\ }\href
  {http://dx.doi.org/10.1103/PhysRevA.99.062309} {\bibfield  {journal}
  {\bibinfo  {journal} {Phys. Rev. A}\ }\textbf {\bibinfo {volume} {99}},\
  \bibinfo {pages} {062309} (\bibinfo {year} {2019})}\BibitemShut {NoStop}%
\bibitem [{\citenamefont {Pollmann}\ \emph {et~al.}(2010)\citenamefont
  {Pollmann}, \citenamefont {Turner}, \citenamefont {Berg},\ and\ \citenamefont
  {Oshikawa}}]{Pollmann:2010aa}%
  \BibitemOpen
  \bibfield  {author} {\bibinfo {author} {\bibfnamefont {F.}~\bibnamefont
  {Pollmann}}, \bibinfo {author} {\bibfnamefont {A.~M.}\ \bibnamefont
  {Turner}}, \bibinfo {author} {\bibfnamefont {E.}~\bibnamefont {Berg}}, \ and\
  \bibinfo {author} {\bibfnamefont {M.}~\bibnamefont {Oshikawa}},\ }\href
  {https://doi.org/10.1103/physrevb.81.064439} {\bibfield  {journal} {\bibinfo
  {journal} {Phys. Rev. B}\ }\textbf {\bibinfo {volume} {81}},\ \bibinfo
  {pages} {064439} (\bibinfo {year} {2010})}\BibitemShut {NoStop}%
\bibitem [{\citenamefont {Fidkowski}(2010)}]{Fidkowski:aa}%
  \BibitemOpen
  \bibfield  {author} {\bibinfo {author} {\bibfnamefont {L.}~\bibnamefont
  {Fidkowski}},\ }\href {https://doi.org/10.1103/physrevlett.104.130502}
  {\bibfield  {journal} {\bibinfo  {journal} {Phys. Rev. Lett.}\ }\textbf
  {\bibinfo {volume} {104}},\ \bibinfo {pages} {130502} (\bibinfo {year}
  {2010})}\BibitemShut {NoStop}%
\bibitem [{\citenamefont {Turner}\ \emph {et~al.}(2011)\citenamefont {Turner},
  \citenamefont {Pollmann},\ and\ \citenamefont {Berg}}]{Turner:2011aa}%
  \BibitemOpen
  \bibfield  {author} {\bibinfo {author} {\bibfnamefont {A.~M.}\ \bibnamefont
  {Turner}}, \bibinfo {author} {\bibfnamefont {F.}~\bibnamefont {Pollmann}}, \
  and\ \bibinfo {author} {\bibfnamefont {E.}~\bibnamefont {Berg}},\ }\href
  {\doibase 10.1103/PhysRevB.83.075102} {\bibfield  {journal} {\bibinfo
  {journal} {Phys. Rev. B}\ }\textbf {\bibinfo {volume} {83}},\ \bibinfo
  {pages} {075102} (\bibinfo {year} {2011})}\BibitemShut {NoStop}%
\bibitem [{Note1()}]{Note1}%
  \BibitemOpen
  \bibinfo {note} {Examples include the equivalence between the entanglement
  spectra of the ground state of finite Ising and Kitaev chains, and spin
  ladders~\cite {Pollmann:2010aa}.}\BibitemShut {Stop}%
\bibitem [{\citenamefont {Katsura}\ \emph {et~al.}(2015)\citenamefont
  {Katsura}, \citenamefont {Schuricht},\ and\ \citenamefont
  {Takahashi}}]{Katsura15}%
  \BibitemOpen
  \bibfield  {author} {\bibinfo {author} {\bibfnamefont {H.}~\bibnamefont
  {Katsura}}, \bibinfo {author} {\bibfnamefont {D.}~\bibnamefont {Schuricht}},
  \ and\ \bibinfo {author} {\bibfnamefont {M.}~\bibnamefont {Takahashi}},\
  }\href {https://doi.org/10.1103/physrevb.92.115137} {\bibfield  {journal}
  {\bibinfo  {journal} {Phys. Rev. B}\ }\textbf {\bibinfo {volume} {92}},\
  \bibinfo {pages} {115137} (\bibinfo {year} {2015})}\BibitemShut {NoStop}%
\bibitem [{\citenamefont {Kitaev}(2001)}]{Kitaev01}%
  \BibitemOpen
  \bibfield  {author} {\bibinfo {author} {\bibfnamefont {A.~Y.}\ \bibnamefont
  {Kitaev}},\ }\href {\doibase 10.1070/1063-7869/44/10s/s29} {\bibfield
  {journal} {\bibinfo  {journal} {Phys.-Usp.}\ }\textbf {\bibinfo {volume}
  {44}},\ \bibinfo {pages} {131} (\bibinfo {year} {2001})}\BibitemShut
  {NoStop}%
\bibitem [{\citenamefont {Beenakker}(2013)}]{Beenakker:2013aa}%
  \BibitemOpen
  \bibfield  {author} {\bibinfo {author} {\bibfnamefont {C.}~\bibnamefont
  {Beenakker}},\ }\href {\doibase 10.1146/annurev-conmatphys-030212-184337}
  {\bibfield  {journal} {\bibinfo  {journal} {Ann. Rev. Cond. Matt. Phys.}\
  }\textbf {\bibinfo {volume} {4}},\ \bibinfo {pages} {113} (\bibinfo {year}
  {2013})}\BibitemShut {NoStop}%
\bibitem [{\citenamefont {Brydges}\ \emph {et~al.}(2019)\citenamefont
  {Brydges}, \citenamefont {Elben}, \citenamefont {Jurcevic}, \citenamefont
  {Vermersch}, \citenamefont {Maier}, \citenamefont {Lanyon}, \citenamefont
  {Zoller}, \citenamefont {Blatt},\ and\ \citenamefont
  {Roos}}]{Brydges:2019aa}%
  \BibitemOpen
  \bibfield  {author} {\bibinfo {author} {\bibfnamefont {T.}~\bibnamefont
  {Brydges}}, \bibinfo {author} {\bibfnamefont {A.}~\bibnamefont {Elben}},
  \bibinfo {author} {\bibfnamefont {P.}~\bibnamefont {Jurcevic}}, \bibinfo
  {author} {\bibfnamefont {B.}~\bibnamefont {Vermersch}}, \bibinfo {author}
  {\bibfnamefont {C.}~\bibnamefont {Maier}}, \bibinfo {author} {\bibfnamefont
  {B.~P.}\ \bibnamefont {Lanyon}}, \bibinfo {author} {\bibfnamefont
  {P.}~\bibnamefont {Zoller}}, \bibinfo {author} {\bibfnamefont
  {R.}~\bibnamefont {Blatt}}, \ and\ \bibinfo {author} {\bibfnamefont {C.~F.}\
  \bibnamefont {Roos}},\ }\href {\doibase 10.1126/science.aau4963} {\bibfield
  {journal} {\bibinfo  {journal} {Science}\ }\textbf {\bibinfo {volume}
  {364}},\ \bibinfo {pages} {260} (\bibinfo {year} {2019})}\BibitemShut
  {NoStop}%
\bibitem [{\citenamefont {van Enk}\ and\ \citenamefont
  {Beenakker}(2012)}]{Enk:2012aa}%
  \BibitemOpen
  \bibfield  {author} {\bibinfo {author} {\bibfnamefont {S.~J.}\ \bibnamefont
  {van Enk}}\ and\ \bibinfo {author} {\bibfnamefont {C.~W.~J.}\ \bibnamefont
  {Beenakker}},\ }\href {https://doi.org/10.1103/PhysRevLett.108.110503}
  {\bibfield  {journal} {\bibinfo  {journal} {Phys.Rev.Lett.}\ }\textbf
  {\bibinfo {volume} {108}},\ \bibinfo {pages} {110503} (\bibinfo {year}
  {2012})}\BibitemShut {NoStop}%
\bibitem [{\citenamefont {Casini}\ and\ \citenamefont
  {Huerta}(2004)}]{Casini:2004ab}%
  \BibitemOpen
  \bibfield  {author} {\bibinfo {author} {\bibfnamefont {H.}~\bibnamefont
  {Casini}}\ and\ \bibinfo {author} {\bibfnamefont {M.}~\bibnamefont
  {Huerta}},\ }\href {\doibase 10.1016/j.physletb.2004.08.072} {\bibfield
  {journal} {\bibinfo  {journal} {Phys. Lett. B}\ }\textbf {\bibinfo {volume}
  {600}},\ \bibinfo {pages} {142} (\bibinfo {year} {2004})}\BibitemShut
  {NoStop}%
\bibitem [{\citenamefont {Wang}\ \emph {et~al.}(2015)\citenamefont {Wang},
  \citenamefont {Xu}, \citenamefont {Wang},\ and\ \citenamefont
  {Wu}}]{Wang2015}%
  \BibitemOpen
  \bibfield  {author} {\bibinfo {author} {\bibfnamefont {D.}~\bibnamefont
  {Wang}}, \bibinfo {author} {\bibfnamefont {S.}~\bibnamefont {Xu}}, \bibinfo
  {author} {\bibfnamefont {Y.}~\bibnamefont {Wang}}, \ and\ \bibinfo {author}
  {\bibfnamefont {C.}~\bibnamefont {Wu}},\ }\href
  {https://doi.org/10.1103/physrevb.91.115118} {\bibfield  {journal} {\bibinfo
  {journal} {Physical Review B}\ }\textbf {\bibinfo {volume} {91}} (\bibinfo
  {year} {2015})}\BibitemShut {NoStop}%
\bibitem [{\citenamefont {Zeng}\ and\ \citenamefont {Zhou}(2016)}]{Zeng:aa}%
  \BibitemOpen
  \bibfield  {author} {\bibinfo {author} {\bibfnamefont {B.}~\bibnamefont
  {Zeng}}\ and\ \bibinfo {author} {\bibfnamefont {D.~L.}\ \bibnamefont
  {Zhou}},\ }\href {\doibase 10.1209/0295-5075/113/56001} {\bibfield  {journal}
  {\bibinfo  {journal} {{EPL}}\ }\textbf {\bibinfo {volume} {113}},\ \bibinfo
  {pages} {56001} (\bibinfo {year} {2016})}\BibitemShut {NoStop}%
\bibitem [{\citenamefont {Kim}(2014)}]{Kim2014}%
  \BibitemOpen
  \bibfield  {author} {\bibinfo {author} {\bibfnamefont {I.~H.}\ \bibnamefont
  {Kim}},\ }\href {https://doi.org/10.1103/physrevb.89.235120} {\bibfield
  {journal} {\bibinfo  {journal} {Physical Review B}\ }\textbf {\bibinfo
  {volume} {89}} (\bibinfo {year} {2014})}\BibitemShut {NoStop}%
\bibitem [{\citenamefont {Stoudenmire}\ \emph {et~al.}(2011)\citenamefont
  {Stoudenmire}, \citenamefont {Alicea}, \citenamefont {Starykh},\ and\
  \citenamefont {Fisher}}]{Stoudenmire11}%
  \BibitemOpen
  \bibfield  {author} {\bibinfo {author} {\bibfnamefont {E.~M.}\ \bibnamefont
  {Stoudenmire}}, \bibinfo {author} {\bibfnamefont {J.}~\bibnamefont {Alicea}},
  \bibinfo {author} {\bibfnamefont {O.~A.}\ \bibnamefont {Starykh}}, \ and\
  \bibinfo {author} {\bibfnamefont {M.~P.}\ \bibnamefont {Fisher}},\ }\href
  {https://doi.org/10.1103/physrevb.84.014503} {\bibfield  {journal} {\bibinfo
  {journal} {Phys. Rev. B}\ }\textbf {\bibinfo {volume} {84}} (\bibinfo {year}
  {2011})}\BibitemShut {NoStop}%
\bibitem [{\citenamefont {Caraglio}\ and\ \citenamefont
  {Gliozzi}(2008)}]{Caraglio:2008aa}%
  \BibitemOpen
  \bibfield  {author} {\bibinfo {author} {\bibfnamefont {M.}~\bibnamefont
  {Caraglio}}\ and\ \bibinfo {author} {\bibfnamefont {F.}~\bibnamefont
  {Gliozzi}},\ }\href {https://doi.org/10.1088/1126-6708/2008/11/076}
  {\bibfield  {journal} {\bibinfo  {journal} {JHEP}\ }\textbf {\bibinfo
  {volume} {0811}},\ \bibinfo {pages} {076} (\bibinfo {year}
  {2008})}\BibitemShut {NoStop}%
\bibitem [{\citenamefont {Furukawa}\ \emph {et~al.}(2009)\citenamefont
  {Furukawa}, \citenamefont {Pasquier},\ and\ \citenamefont
  {Shiraishi}}]{Furukawa:2009aa}%
  \BibitemOpen
  \bibfield  {author} {\bibinfo {author} {\bibfnamefont {S.}~\bibnamefont
  {Furukawa}}, \bibinfo {author} {\bibfnamefont {V.}~\bibnamefont {Pasquier}},
  \ and\ \bibinfo {author} {\bibfnamefont {J.}~\bibnamefont {Shiraishi}},\
  }\href {https://doi.org/10.1103/PhysRevLett.102.170602} {\bibfield  {journal}
  {\bibinfo  {journal} {Phys. Rev. Lett.}\ }\textbf {\bibinfo {volume} {102}},\
  \bibinfo {pages} {170602} (\bibinfo {year} {2009})}\BibitemShut {NoStop}%
\bibitem [{\citenamefont {McCoy}\ and\ \citenamefont
  {Yan}(1983)}]{MCCOY1983278}%
  \BibitemOpen
  \bibfield  {author} {\bibinfo {author} {\bibfnamefont {B.~M.}\ \bibnamefont
  {McCoy}}\ and\ \bibinfo {author} {\bibfnamefont {M.-L.}\ \bibnamefont
  {Yan}},\ }\href {\doibase https://doi.org/10.1016/0550-3213(83)90216-X}
  {\bibfield  {journal} {\bibinfo  {journal} {Nuc. Phys. B}\ }\textbf {\bibinfo
  {volume} {215}},\ \bibinfo {pages} {278 } (\bibinfo {year}
  {1983})}\BibitemShut {NoStop}%
\bibitem [{\citenamefont {Lacroix}\ \emph {et~al.}(2010)\citenamefont
  {Lacroix}, \citenamefont {Mendels},\ and\ \citenamefont
  {Mila}}]{Lacroix2010}%
  \BibitemOpen
  \bibinfo {editor} {\bibfnamefont {C.}~\bibnamefont {Lacroix}}, \bibinfo
  {editor} {\bibfnamefont {P.}~\bibnamefont {Mendels}}, \ and\ \bibinfo
  {editor} {\bibfnamefont {F.}~\bibnamefont {Mila}},\ eds.,\ \href@noop {}
  {\emph {\bibinfo {title} {Introduction to Frustrated Magnetism}}}\ (\bibinfo
  {publisher} {Springer Series in Solid-State Sciences Vol. 164},\ \bibinfo
  {year} {2010})\BibitemShut {NoStop}%
\bibitem [{\citenamefont {Zeng}\ \emph {et~al.}(2019)\citenamefont {Zeng},
  \citenamefont {Chen}, \citenamefont {Zhou},\ and\ \citenamefont
  {Wen}}]{Zeng19}%
  \BibitemOpen
  \bibfield  {author} {\bibinfo {author} {\bibfnamefont {B.}~\bibnamefont
  {Zeng}}, \bibinfo {author} {\bibfnamefont {X.}~\bibnamefont {Chen}}, \bibinfo
  {author} {\bibfnamefont {D.-L.}\ \bibnamefont {Zhou}}, \ and\ \bibinfo
  {author} {\bibfnamefont {X.-G.}\ \bibnamefont {Wen}},\ }\href {\doibase
  10.1007/978-1-4939-9084-9} {\emph {\bibinfo {title} {Quantum Information
  Meets Quantum Matter}}}\ (\bibinfo  {publisher} {Springer New York},\
  \bibinfo {year} {2019})\BibitemShut {NoStop}%
\bibitem [{\citenamefont {Peschel}(2003{\natexlab{a}})}]{Peshel03}%
  \BibitemOpen
  \bibfield  {author} {\bibinfo {author} {\bibfnamefont {I.}~\bibnamefont
  {Peschel}},\ }\href {\doibase 10.1088/0305-4470/36/14/101} {\bibfield
  {journal} {\bibinfo  {journal} {Jour. Phys. A: Math. Gen.}\ }\textbf
  {\bibinfo {volume} {36}},\ \bibinfo {pages} {L205} (\bibinfo {year}
  {2003}{\natexlab{a}})}\BibitemShut {NoStop}%
\bibitem [{\citenamefont {White}(1992)}]{White1992}%
  \BibitemOpen
  \bibfield  {author} {\bibinfo {author} {\bibfnamefont {S.~R.}\ \bibnamefont
  {White}},\ }\href {\doibase 10.1103/PhysRevLett.69.2863} {\bibfield
  {journal} {\bibinfo  {journal} {Phys. Rev. Lett.}\ }\textbf {\bibinfo
  {volume} {69}},\ \bibinfo {pages} {2863} (\bibinfo {year}
  {1992})}\BibitemShut {NoStop}%
\bibitem [{\citenamefont {Schollw{\"{o}}ck}(2005)}]{Schollwock2005}%
  \BibitemOpen
  \bibfield  {author} {\bibinfo {author} {\bibfnamefont {U.}~\bibnamefont
  {Schollw{\"{o}}ck}},\ }\href {\doibase 10.1103/RevModPhys.77.259} {\bibfield
  {journal} {\bibinfo  {journal} {Rev. Mod. Phys.}\ }\textbf {\bibinfo {volume}
  {77}},\ \bibinfo {pages} {259} (\bibinfo {year} {2005})}\BibitemShut
  {NoStop}%
\bibitem [{\citenamefont {Caio}\ \emph {et~al.}(2015)\citenamefont {Caio},
  \citenamefont {Cooper},\ and\ \citenamefont {Bhaseen}}]{Caio:2015aa}%
  \BibitemOpen
  \bibfield  {author} {\bibinfo {author} {\bibfnamefont {M.~D.}\ \bibnamefont
  {Caio}}, \bibinfo {author} {\bibfnamefont {N.~R.}\ \bibnamefont {Cooper}}, \
  and\ \bibinfo {author} {\bibfnamefont {M.~J.}\ \bibnamefont {Bhaseen}},\
  }\href@noop {} {\bibfield  {journal} {\bibinfo  {journal} {Phys. Rev. Lett.}\
  }\textbf {\bibinfo {volume} {115}},\ \bibinfo {pages} {236403} (\bibinfo
  {year} {2015})}\BibitemShut {NoStop}%
\bibitem [{\citenamefont {D'Alessio}\ and\ \citenamefont
  {Rigol}(2015)}]{DAlessio:2015aa}%
  \BibitemOpen
  \bibfield  {author} {\bibinfo {author} {\bibfnamefont {L.}~\bibnamefont
  {D'Alessio}}\ and\ \bibinfo {author} {\bibfnamefont {M.}~\bibnamefont
  {Rigol}},\ }\href@noop {} {\bibfield  {journal} {\bibinfo  {journal} {Nat.
  Commun.}\ }\textbf {\bibinfo {volume} {6}},\ \bibinfo {pages} {8336}
  (\bibinfo {year} {2015})}\BibitemShut {NoStop}%
\bibitem [{\citenamefont {McGinley}\ and\ \citenamefont
  {Cooper}()}]{McGinley:aa}%
  \BibitemOpen
  \bibfield  {author} {\bibinfo {author} {\bibfnamefont {M.}~\bibnamefont
  {McGinley}}\ and\ \bibinfo {author} {\bibfnamefont {N.~R.}\ \bibnamefont
  {Cooper}},\ }\href@noop {} {\bibinfo  {journal} {arXiv:1908.06875}\
  }\BibitemShut {NoStop}%
\bibitem [{\citenamefont {Giamarchi}(2004)}]{Giamarchi2004}%
  \BibitemOpen
\bibfield  {journal} {  }\bibfield  {author} {\bibinfo {author} {\bibfnamefont
  {T.}~\bibnamefont {Giamarchi}},\ }\href {\doibase
  10.1093/acprof:oso/9780198525004.001.0001} {\emph {\bibinfo {title} {{Quantum
  physics in one dimension}}}},\ Internat. Ser. Mono. Phys.\ (\bibinfo
  {publisher} {Clarendon Press},\ \bibinfo {address} {Oxford},\ \bibinfo {year}
  {2004})\BibitemShut {NoStop}%
\bibitem [{\citenamefont {Gergs}\ \emph {et~al.}(2016)\citenamefont {Gergs},
  \citenamefont {Fritz},\ and\ \citenamefont {Schuricht}}]{Gergs2016}%
  \BibitemOpen
  \bibfield  {author} {\bibinfo {author} {\bibfnamefont {N.~M.}\ \bibnamefont
  {Gergs}}, \bibinfo {author} {\bibfnamefont {L.}~\bibnamefont {Fritz}}, \ and\
  \bibinfo {author} {\bibfnamefont {D.}~\bibnamefont {Schuricht}},\ }\href
  {https://doi.org/10.1103/physrevb.93.075129} {\bibfield  {journal} {\bibinfo
  {journal} {Physical Review B}\ }\textbf {\bibinfo {volume} {93}} (\bibinfo
  {year} {2016})}\BibitemShut {NoStop}%
\bibitem [{\citenamefont {Levy}\ and\ \citenamefont
  {Goldstein}(2019)}]{Levy2019}%
  \BibitemOpen
  \bibfield  {author} {\bibinfo {author} {\bibfnamefont {L.}~\bibnamefont
  {Levy}}\ and\ \bibinfo {author} {\bibfnamefont {M.}~\bibnamefont
  {Goldstein}},\ }\href {\doibase 10.3390/universe5010033} {\bibfield
  {journal} {\bibinfo  {journal} {Universe}\ }\textbf {\bibinfo {volume} {5}},\
  \bibinfo {pages} {33} (\bibinfo {year} {2019})}\BibitemShut {NoStop}%
\bibitem [{\citenamefont {Niu}\ \emph {et~al.}(1985)\citenamefont {Niu},
  \citenamefont {Thouless},\ and\ \citenamefont {Wu}}]{Niu1985}%
  \BibitemOpen
  \bibfield  {author} {\bibinfo {author} {\bibfnamefont {Q.}~\bibnamefont
  {Niu}}, \bibinfo {author} {\bibfnamefont {D.~J.}\ \bibnamefont {Thouless}}, \
  and\ \bibinfo {author} {\bibfnamefont {Y.-S.}\ \bibnamefont {Wu}},\ }\href
  {\doibase 10.1103/physrevb.31.3372} {\bibfield  {journal} {\bibinfo
  {journal} {Phys. Rev. B}\ }\textbf {\bibinfo {volume} {31}},\ \bibinfo
  {pages} {3372} (\bibinfo {year} {1985})}\BibitemShut {NoStop}%
\bibitem [{\citenamefont {Chen}\ \emph {et~al.}(2012)\citenamefont {Chen},
  \citenamefont {Gu}, \citenamefont {Liu},\ and\ \citenamefont
  {Wen}}]{Chen:2012aa}%
  \BibitemOpen
  \bibfield  {author} {\bibinfo {author} {\bibfnamefont {X.}~\bibnamefont
  {Chen}}, \bibinfo {author} {\bibfnamefont {Z.-C.}\ \bibnamefont {Gu}},
  \bibinfo {author} {\bibfnamefont {Z.-X.}\ \bibnamefont {Liu}}, \ and\
  \bibinfo {author} {\bibfnamefont {X.-G.}\ \bibnamefont {Wen}},\ }\href
  {https://science.sciencemag.org/content/338/6114/1604} {\bibfield  {journal}
  {\bibinfo  {journal} {Science}\ }\textbf {\bibinfo {volume} {338}},\ \bibinfo
  {pages} {1604} (\bibinfo {year} {2012})}\BibitemShut {NoStop}%
\bibitem [{\citenamefont {Haegeman}\ \emph {et~al.}(2012)\citenamefont
  {Haegeman}, \citenamefont {Perez-Garcia}, \citenamefont {Cirac},\ and\
  \citenamefont {Schuch}}]{Haegeman:2012aa}%
  \BibitemOpen
  \bibfield  {author} {\bibinfo {author} {\bibfnamefont {J.}~\bibnamefont
  {Haegeman}}, \bibinfo {author} {\bibfnamefont {D.}~\bibnamefont
  {Perez-Garcia}}, \bibinfo {author} {\bibfnamefont {I.}~\bibnamefont {Cirac}},
  \ and\ \bibinfo {author} {\bibfnamefont {N.}~\bibnamefont {Schuch}},\
  }\href@noop {} {\bibfield  {journal} {\bibinfo  {journal} {Phys. Rev. Lett.}\
  }\textbf {\bibinfo {volume} {109}},\ \bibinfo {pages} {050402} (\bibinfo
  {year} {2012})}\BibitemShut {NoStop}%
\bibitem [{\citenamefont {Pollmann}\ and\ \citenamefont
  {Turner}(2012)}]{Pollmann:2012aa}%
  \BibitemOpen
  \bibfield  {author} {\bibinfo {author} {\bibfnamefont {F.}~\bibnamefont
  {Pollmann}}\ and\ \bibinfo {author} {\bibfnamefont {A.~M.}\ \bibnamefont
  {Turner}},\ }\href@noop {} {\bibfield  {journal} {\bibinfo  {journal} {Phys.
  Rev. B}\ }\textbf {\bibinfo {volume} {86}},\ \bibinfo {pages} {125441}
  (\bibinfo {year} {2012})}\BibitemShut {NoStop}%
\bibitem [{\citenamefont {Shapourian}\ \emph
  {et~al.}(2017{\natexlab{a}})\citenamefont {Shapourian}, \citenamefont
  {Shiozaki},\ and\ \citenamefont {Ryu}}]{Shapourian2017b}%
  \BibitemOpen
  \bibfield  {author} {\bibinfo {author} {\bibfnamefont {H.}~\bibnamefont
  {Shapourian}}, \bibinfo {author} {\bibfnamefont {K.}~\bibnamefont
  {Shiozaki}}, \ and\ \bibinfo {author} {\bibfnamefont {S.}~\bibnamefont
  {Ryu}},\ }\href {\doibase 10.1103/PhysRevLett.118.216402} {\bibfield
  {journal} {\bibinfo  {journal} {Phys. Rev. Lett.}\ }\textbf {\bibinfo
  {volume} {118}},\ \bibinfo {pages} {216402} (\bibinfo {year}
  {2017}{\natexlab{a}})}\BibitemShut {NoStop}%
\bibitem [{\citenamefont {Shapourian}\ \emph
  {et~al.}(2017{\natexlab{b}})\citenamefont {Shapourian}, \citenamefont
  {Shiozaki},\ and\ \citenamefont {Ryu}}]{Shapourian2017a}%
  \BibitemOpen
  \bibfield  {author} {\bibinfo {author} {\bibfnamefont {H.}~\bibnamefont
  {Shapourian}}, \bibinfo {author} {\bibfnamefont {K.}~\bibnamefont
  {Shiozaki}}, \ and\ \bibinfo {author} {\bibfnamefont {S.}~\bibnamefont
  {Ryu}},\ }\href {\doibase 10.1103/PhysRevB.95.165101} {\bibfield  {journal}
  {\bibinfo  {journal} {Phys. Rev. B}\ }\textbf {\bibinfo {volume} {95}},\
  \bibinfo {pages} {165101} (\bibinfo {year} {2017}{\natexlab{b}})}\BibitemShut
  {NoStop}%
\bibitem [{\citenamefont {Elben}\ \emph {et~al.}(2019)\citenamefont {Elben},
  \citenamefont {Yu}, \citenamefont {Zhu}, \citenamefont {Hafezi},
  \citenamefont {Pollmann}, \citenamefont {Zoller},\ and\ \citenamefont
  {Vermersch}}]{Elben:aa}%
  \BibitemOpen
  \bibfield  {author} {\bibinfo {author} {\bibfnamefont {A.}~\bibnamefont
  {Elben}}, \bibinfo {author} {\bibfnamefont {J.}~\bibnamefont {Yu}}, \bibinfo
  {author} {\bibfnamefont {G.}~\bibnamefont {Zhu}}, \bibinfo {author}
  {\bibfnamefont {M.}~\bibnamefont {Hafezi}}, \bibinfo {author} {\bibfnamefont
  {F.}~\bibnamefont {Pollmann}}, \bibinfo {author} {\bibfnamefont
  {P.}~\bibnamefont {Zoller}}, \ and\ \bibinfo {author} {\bibfnamefont
  {B.}~\bibnamefont {Vermersch}},\ }\href {https://arxiv.org/abs/1906.05011}
  {\bibfield  {journal} {\bibinfo  {journal} {arXiv:1906.05011}\ } (\bibinfo
  {year} {2019})}\BibitemShut {NoStop}%
\bibitem [{\citenamefont {Ollivier}\ and\ \citenamefont
  {Zurek}(2001)}]{Ollivier:2001aa}%
  \BibitemOpen
  \bibfield  {author} {\bibinfo {author} {\bibfnamefont {H.}~\bibnamefont
  {Ollivier}}\ and\ \bibinfo {author} {\bibfnamefont {W.~H.}\ \bibnamefont
  {Zurek}},\ }\href {\doibase 10.1103/PhysRevLett.88.017901} {\bibfield
  {journal} {\bibinfo  {journal} {Phys. Revs Lett.}\ }\textbf {\bibinfo
  {volume} {88}},\ \bibinfo {pages} {017901} (\bibinfo {year}
  {2001})}\BibitemShut {NoStop}%
\bibitem [{\citenamefont {Fr{\'e}rot}\ and\ \citenamefont
  {Roscilde}(2016)}]{Frerot:2016aa}%
  \BibitemOpen
  \bibfield  {author} {\bibinfo {author} {\bibfnamefont {I.}~\bibnamefont
  {Fr{\'e}rot}}\ and\ \bibinfo {author} {\bibfnamefont {T.}~\bibnamefont
  {Roscilde}},\ }\href {https://doi.org/10.1103/PhysRevB.94.075121} {\bibfield
  {journal} {\bibinfo  {journal} {Phys. Rev. B}\ }\textbf {\bibinfo {volume}
  {94}},\ \bibinfo {pages} {075121} (\bibinfo {year} {2016})}\BibitemShut
  {NoStop}%
\bibitem [{\citenamefont {Peschel}\ and\ \citenamefont
  {Chung}(1999)}]{Peschel01}%
  \BibitemOpen
  \bibfield  {author} {\bibinfo {author} {\bibfnamefont {I.}~\bibnamefont
  {Peschel}}\ and\ \bibinfo {author} {\bibfnamefont {M.-C.}\ \bibnamefont
  {Chung}},\ }\href {\doibase 10.1088/0305-4470/32/48/305} {\bibfield
  {journal} {\bibinfo  {journal} {Jour. Phys. A: Math. Gen.}\ }\textbf
  {\bibinfo {volume} {32}},\ \bibinfo {pages} {8419} (\bibinfo {year}
  {1999})}\BibitemShut {NoStop}%
\bibitem [{\citenamefont {Chung}\ and\ \citenamefont
  {Peschel}(2000)}]{Peschel02}%
  \BibitemOpen
  \bibfield  {author} {\bibinfo {author} {\bibfnamefont {M.-C.}\ \bibnamefont
  {Chung}}\ and\ \bibinfo {author} {\bibfnamefont {I.}~\bibnamefont
  {Peschel}},\ }\href {\doibase 10.1103/PhysRevB.62.4191} {\bibfield  {journal}
  {\bibinfo  {journal} {Phys. Rev. B}\ }\textbf {\bibinfo {volume} {62}},\
  \bibinfo {pages} {4191} (\bibinfo {year} {2000})}\BibitemShut {NoStop}%
\bibitem [{\citenamefont {Chung}\ and\ \citenamefont
  {Peschel}(2001)}]{Peschel03}%
  \BibitemOpen
  \bibfield  {author} {\bibinfo {author} {\bibfnamefont {M.-C.}\ \bibnamefont
  {Chung}}\ and\ \bibinfo {author} {\bibfnamefont {I.}~\bibnamefont
  {Peschel}},\ }\href {\doibase 10.1103/PhysRevB.64.064412} {\bibfield
  {journal} {\bibinfo  {journal} {Phys. Rev. B}\ }\textbf {\bibinfo {volume}
  {64}},\ \bibinfo {pages} {064412} (\bibinfo {year} {2001})}\BibitemShut
  {NoStop}%
\bibitem [{\citenamefont {Peschel}(2003{\natexlab{b}})}]{Peschel04}%
  \BibitemOpen
  \bibfield  {author} {\bibinfo {author} {\bibfnamefont {I.}~\bibnamefont
  {Peschel}},\ }\href {\doibase 10.1088/0305-4470/36/14/101} {\bibfield
  {journal} {\bibinfo  {journal} {Journal of Physics A: Mathematical and
  General}\ }\textbf {\bibinfo {volume} {36}},\ \bibinfo {pages} {L205}
  (\bibinfo {year} {2003}{\natexlab{b}})}\BibitemShut {NoStop}%
\bibitem [{mpm()}]{mpmath_library}%
  \BibitemOpen
  \href@noop {} {}\bibinfo {note} {Fredrik Johansson and others. mpmath: a
  Python library for arbitrary-precision floating-point arithmetic,
  http://mpmath.org/.}\BibitemShut {Stop}%
\bibitem [{\citenamefont {Peschel}\ and\ \citenamefont
  {Eisler}(2009)}]{Peschel09}%
  \BibitemOpen
  \bibfield  {author} {\bibinfo {author} {\bibfnamefont {I.}~\bibnamefont
  {Peschel}}\ and\ \bibinfo {author} {\bibfnamefont {V.}~\bibnamefont
  {Eisler}},\ }\href {\doibase 10.1088/1751-8113/42/50/504003} {\bibfield
  {journal} {\bibinfo  {journal} {Jour. Phys. A: Math. Theo.}\ }\textbf
  {\bibinfo {volume} {42}},\ \bibinfo {pages} {504003} (\bibinfo {year}
  {2009})}\BibitemShut {NoStop}%
\bibitem [{\citenamefont {Fagotti}\ and\ \citenamefont
  {Calabrese}(2008)}]{Fagotti08}%
  \BibitemOpen
  \bibfield  {author} {\bibinfo {author} {\bibfnamefont {M.}~\bibnamefont
  {Fagotti}}\ and\ \bibinfo {author} {\bibfnamefont {P.}~\bibnamefont
  {Calabrese}},\ }\href {\doibase 10.1103/PhysRevA.78.010306} {\bibfield
  {journal} {\bibinfo  {journal} {Phys. Rev. A}\ }\textbf {\bibinfo {volume}
  {78}},\ \bibinfo {pages} {010306} (\bibinfo {year} {2008})}\BibitemShut
  {NoStop}%
\bibitem [{\citenamefont {Schollw{\"{o}}ck}(2011)}]{DMRG_Schollwoeck}%
  \BibitemOpen
  \bibfield  {author} {\bibinfo {author} {\bibfnamefont {U.}~\bibnamefont
  {Schollw{\"{o}}ck}},\ }\href {\doibase
  https://doi.org/10.1016/j.aop.2010.09.012} {\bibfield  {journal} {\bibinfo
  {journal} {Ann. of Phys.}\ }\textbf {\bibinfo {volume} {326}},\ \bibinfo
  {pages} {96 } (\bibinfo {year} {2011})}\BibitemShut {NoStop}%
\end{thebibliography}%

%%%%%%%%%%%%%%%%%%%%%%%%%%%%%%%%%%%%%%%%%%%%%%%%%%%%%%%%%
\end{document}